\documentclass{ieeeaccess_mod}
\usepackage{cite}
\usepackage{amsmath,amssymb,amsfonts}
\usepackage{algorithmic}
\usepackage{graphicx}
\usepackage{textcomp}
\usepackage[useregional]{datetime2}
\usepackage{subfig}
\usepackage{float}
\usepackage{multirow}
\usepackage{array}
\usepackage[dvipsnames]{xcolor}

\graphicspath{{./figures/}{./}}

\def\BibTeX{{\rm B\kern-.05em{\sc i\kern-.025em b}\kern-.08em
    T\kern-.1667em\lower.7ex\hbox{E}\kern-.125emX}}
\begin{document}
\history{}
\doi{}

\title{Semi-supervised source localization in reverberant environments with deep generative modeling}
\author{\uppercase{Michael J. Bianco}\authorrefmark{1}, \IEEEmembership{Member, IEEE},
\uppercase{Sharon Gannot\authorrefmark{2}}, \IEEEmembership{Fellow, IEEE}, 
\uppercase{Efren Fernandez-Grande}\authorrefmark{3}, \IEEEmembership{Member, IEEE}
and \uppercase{Peter Gerstoft}\authorrefmark{1}, \IEEEmembership{Senior Member, IEEE}}
\address[1]{Marine Physical Laboratory, University of California San Diego, La Jolla, CA 92093, USA}
\address[2]{Faculty of Engineering, Bar-Ilan University, Ramat-Gan 5290002, Israel}
\address[3]{Department of Electrical Engineering, Technical University Denmark, Kongens Lyngby, Denmark}
\tfootnote{This work is supported by the Office of Naval Research, Grant No. N00014-11-1-0439 and the European Union’s Horizon 2020 Research and Innovation Programme under Grant Agreement No. 871245.}

\markboth
{Bianco \headeretal: Semi-supervised source localization}
{Bianco \headeretal: Semi-supervised source localization}
%{Author \headeretal: Preparation of Papers for IEEE TRANSACTIONS and JOURNALS}

\corresp{Corresponding author: Michael J. Bianco (e-mail: mbianco@ucsd.edu)}

\begin{abstract}
We propose a semi-supervised approach to acoustic source localization in reverberant environments based on deep generative modeling. Localization in reverberant environments remains an open challenge. Even with large data volumes, the number of labels available for supervised learning in reverberant environments is usually small. We address this issue by performing semi-supervised learning (SSL) with convolutional variational autoencoders (VAEs) on reverberant speech signals recorded with microphone arrays. The VAE is trained to generate the phase of relative transfer functions (RTFs) between microphones, in parallel with a direction of arrival (DOA) classifier based on RTF-phase. These models are trained using both labeled and unlabeled RTF-phase sequences. In learning to perform these tasks, the VAE-SSL explicitly learns to separate the physical causes of the RTF-phase (i.e., source location) from distracting signal characteristics such as noise and speech activity. Relative to existing semi-supervised localization methods in acoustics, VAE-SSL is effectively an {\it end-to-end} processing approach which relies on minimal preprocessing of RTF-phase features. As far as we are aware, our paper presents the first approach to modeling the physics of acoustic propagation using deep generative modeling. The VAE-SSL approach is compared with two signal processing-based approaches, steered response power with phase transform (SRP-PHAT) and MUltiple SIgnal Classification (MUSIC), as well as fully supervised CNNs. We find that VAE-SSL can outperform the conventional approaches and the CNN in label-limited scenarios.  Further,  the trained VAE-SSL system can  generate new RTF-phase samples, which shows the  VAE-SSL approach learns the physics of the acoustic environment. The generative modeling in VAE-SSL thus provides a means of interpreting the learned representations. 
\end{abstract}


\titlepgskip=-15pt

\maketitle
% \noindent {\color{blue} \bf DRAFT DATE: \today}

\section{Introduction}
\label{sec:intro}

Source localization is an important problem in acoustics and many related fields. The performance of localization algorithms is degraded by reverberation, which induces complex temporal arrival structure at sensor arrays. Despite recent advances, e.g. \cite{purwins_deep_2019,bianco2019machine,gannot2019introduction}, acoustic localization in reverberant environments remains a major challenge \cite{traer2016statistics}. 
There has been great interest in machine learning (ML)-based techniques  in acoustics, including source localization and event detection \cite{nakashima20053d,deleforge20122d,mesaros2017dcase,vincent2018audio,chakrabarty2019multi,adavanne2019sound,ping2020three,ozanich2020feedforward,zhu2020feature,hammer2020fcn}. One difficulty for ML-based methods in acoustics is the limited amount of labeled data and the complex acoustic propagation in natural environments, despite  large volumes of recordings\cite{purwins_deep_2019,bianco2019machine}. This limitation has motivated recent approaches for localization based on semi-supervised learning (SSL)\cite{laufer2016semi,opochinsky2019deep}. 

We approach source localization from the perspective of SSL, with the intent of addressing real-world applications of ML. It has been shown that it is relatively easy to generate large amounts of synthetic data resembling real-world sound measurement configurations. This synthetic data has then been used to train ML-based localization models (e.g.\cite{chakrabarty2019multi}), with very good performance. However, in most real scenarios, room geometry includes irregular boundaries, scattering, and diffracting elements (e.g., furniture and uneven surfaces) which may not be convenient to model using acoustic propagation software.

In this paper we contend, along the SSL paradigm, that if there is a microphone system with fixed geometry, recording in a room environment, that reverberant signals convey physical characteristics of the room - that given sufficient time and variety of source locations, the physics of the room may be well modeled using unsupervised ML \cite{gannot2001signal,laufer2016semi,gannot2017consolidated} . We further observe that if few labels are available, obtained for instance using cell-phone data, we can leverage such unsupervised representations to localized sources in the room.

We propose an SSL localization approach based on deep generative modeling with variational autoencoders (VAEs)\cite{kingma2014auto,kingma2014semi,kingma2019introduction}. Deep generative models \cite{goodfellow2016deep}, e.g. generative adversarial networks (GANs)\cite{goodfellow2014generative}, have received much attention for their ability to learn high-dimensional data distributions, including those of natural images \cite{karras2019analyzing}. GANs in acoustics have had success in generating raw audio \cite{oord2016wavenet} and speech enhancement \cite{donahue2018exploring}. VAEs are inspiring examples of representation learning \cite{bengio2013representation} which use both encoder and decoder neural networks (NNs) to parameterize probability distributions. In this way explicit latent codes are obtained for their generative models.

We use VAEs to perform SSL. Our work is based on seminal work in SSL with deep generative models \cite{kingma2014semi}. The approach outlined in \cite{kingma2014semi} has become a well-established method in literature for SSL with high-dimensional data. We extend the philosophy from \cite{kingma2014semi}, which focused on image processing, to the fields of acoustics and array processing. The VAE system is used to model physical data in real world acoustic environments. As far as we are aware, our paper presents the first approach to modeling the physics of acoustic propagation using deep generative modeling.

 In our proposed approach, VAEs are used to encode and generate the phase of the relative transfer function (RTF) between two microphones \cite{gannot2001signal}. The VAE is trained in parallel with a classifier network to benefit from both labeled and unlabeled examples. The resulting model estimates DOA and generates RTF-phase sequences. By learning to generate RTF phase, the VAE-SSL system learns the physical model relating the latent model and DOA label to the RTF-phase. 

This approach is a form of manifold learning \cite{goodfellow2016deep,peterfreund2020local,laufer2020data}. Manifold learning has recently been proposed for semi-supervised source localization in room acoustics \cite{laufer2016semi}. In this work, diffusion mapping is used to obtain a graphical model relating source location to RTF features. It was shown that a lower-dimensional latent representation of the RTFs, extracted by diffusion mapping, correlated well with source positions.

We present our VAE-SSL method as an alternative to this manifold learning approach. Recent work has indicated the capabilities of deep learning in manifold learning\cite{peterfreund2020local}. VAE-SSL uses the non-linear modeling capabilities of deep generative modeling to obtain an SSL localization approach which is less reliant on preprocessing and hand-engineering of latent representations. Thus, the approach can be regarded as nearly {\it end-to-end}. The VAE-SSL system is designed to not rely on  significant preprocessing of the RTFs. Through gradient-based learning it automatically determines the best latent and discriminative representations for the given task. Instead of spectral averaging, we input to the system a sequence of instantaneous RTFs and allow the statistical model to best utilize the patterns in the data.

The generative modeling approach has in general three benefits. First, the system learns a latent embedding of the RTF-phase sequences from the unlabeled data. This representation is leveraged by training with labeled data, which give the latent embedding semantic meaning - in our case, the relationship between the learned embedding and physical source locations in a room. Second, the system allows disentangling of the causes of RTF-phase signal (i.e., source DOA) from other signal factors including source frequency variation \cite{siddharth2017learning,kingma2019introduction}. The disentanglement enables the system to learn a task-specific representation of the RTF-phase patterns in input features that are most relevant to DOA estimation task. Finally, the system can be used to conditionally generate new RTF-phase sequences, based on sampling over the latent representation and DOA label. This output can be interpreted physically, and thereby allows us to verify that the system is obtaining a physically meaningful representation.

% VAE-SSL separates the effect of the label from other signal causes (e.g. noise, and speech dynamics). This focuses the classifier on features of the phase sequence which are relevant to the localization task. 

% Our generative modeling approach disentangles the causes of the RTF-phase (i.e., source DOA) from other signal factors including source frequency variation \cite{siddharth2017learning,kingma2019introduction}. This is accomplished by training in parallel both a generative model and classifier network. The disentanglement enables the system to learn a task-specific representation of the RTF-phase patterns in input features that are most relevant to DOA estimation task. Further, the trained VAE in VAE-SSL can be used to conditionally generate RTF-phase sequences based on sampling over the latent representation and DOA label. This output can be interpreted physically, and thereby allows us to verify that the system is obtaining a physically meaningful representation.

We build upon our previous work in deep generative modeling for source localization \cite{bianco2020semi}. In this work,  the VAE-SSL approach could well learn to localize acoustic sources with a few labeled RTF-phase sequences and extensive unlabeled data. The system was trained and tested on noise signals in simulated environments. 

In our current study, we extend our VAE-SSL concepts
to more realistic acoustic scenarios, refine the inference and generative architectures, and demonstrate and characterize the implicit acoustic model learned by the generative model by using the trained VAE-SSL system to conditionally generate RTF-phase sequences. This includes consideration for the appropriate conditional distributions for the generative model. We train and validate the learning-based approaches on speech data in two reverberant environments. As part of our study, we have obtained a new acoustic dataset at the Technical University of Denmark (DTU). This acoustic dataset consists of reverberant acoustic impulse responses (IRs) recorded in a classroom at DTU from several source locations. In this dataset, we also considered off-grid and off-range measurements to test the generalization of the learning-based methods.

We use speech data from the LibriSpeech corpus \cite{panayotov2015librispeech}. Reverberant speech is obtained by convolving dry speech with estimated IRs from two real-world datasets: the Multichannel Impulse Response (MIR) database \cite{hadad2014multichannel} and the DTU room acoustics dataset.

The VAE-SSL method is implemented using convolutional neural networks (CNNs). The performance of the convolutional VAE-SSL in reverberant environments is assessed against two signal processing-based approaches: the steered response power with phase transform (SRP-PHAT)\cite{brandstein1997robust} and MUltiple SIgnal Classification (MUSIC) \cite{schmidt1986multiple} approaches. We further consider the performance of a fully-supervised convolutional CNN. Comparison of MUSIC and SRP-PHAT in ML DOA estimation  literature, as well as DOA estimation techniques more generally,  is typical, e.g. \cite{chakrabarty2019multi}. Moreover, the MUSIC method was selected as the baseline method in the recent IEEE Signal Processing Society acoustic source LOCalization And TrAcking (LOCATA) challenge\cite{evers2020locata}.

We show that VAE-SSL can outperform conventional source localization approaches, as well as fully supervised approaches for label-limited scenarios. This includes scenarios where the source may be off-range or off-grid from the design case, and also variations in reverberation time and speech signals. We further show the implicit physical model obtained with VAE-SSL can be used to generate RTF-phase features. The physical characteristics of the generated RTF-phase is assessed by analyzing the phase wrap of the generated phase and the corresponding phase-delay for the generated RTF time domain representation.

\section{Theory}
\label{sec:theory}
We use RTFs\cite{gannot2001signal}, specifically the RTF-phase, as the acoustic feature for our VAE-SSL approach. Since the RTF can be assumed independent of the source, this feature helps to focus ML on physically relevant features, and thereby reduces the sample complexity of the model. We encode a temporal sequence of instantaneous RTF-phases as a function of source azimuth (direction of arrival, DOA). We choose the instantaneous RTF-phase, calculated using a single STFT frame (i.e. no averaging applied), to minimize the intervention of feature preprocessing. This allows the VAE-SSL method to extract end-to-end the important features for source localization and RTF-phase generation. This formulation may facilitate the localization of moving sources, namely speaker tracking problems. This extension is left for future work.

\subsection{Relative transfer function (RTF)} 
\label{sec:rtfs}
We consider short-time Fourier transform (STFT) domain acoustic recordings of the form
\begin{align}\label{eq:mics}
d_i(k)&=a_i(k)s(k)+u_i(k),
\end{align}
with $s$  the source signal $a_i$ the acoustic transfer function relating the source and each of the microphones ($i=\{1,2\}$ the microphone index and $k$ the frequency index), and $u_i$ sensor noise (spatially-white).
Then, the relative transfer function (RTF) is defined as\cite{laufer2016semi,gannot2001signal}
\begin{equation}
    h(k) = \frac{a_2(k)}{a_1(k)},
\end{equation}
with $k$ the frequency index. With $d_1$ as reference, the instantaneous RTF $\widehat{h}(k)$ is calculated using a single STFT frame
% % 
% \begin{align}\label{eq:mics_rtf_fft2}
% \widehat{H}(k)=\frac{A_2(k)}{A_1(k)}=\frac{S_{d_2d_1}}{S_{d_1d_1}}, 
% \end{align}
% %
%
\begin{align}\label{eq:mics_rtf_fft2}
\widehat{h}(k)=\frac{d_2(k)}{d_1(k)}.
\end{align}
%
% with $\cdot^{*}$  the complex conjugate. 
This estimator is biased since we neglect the noise spectra [Ref.~\cite{laufer2016semi}, Eq.~(5)]. While an unbiased estimator can be obtained, see \cite{markovich2018performance,koldovsky2015spatial}, we observe that the biased estimate $\widehat{h}(k)$ works well for our purposes. For each STFT frame, a vector of RTFs is obtained $\widehat{\mathbf{h}}=[\widehat{h}(1)\dots \widehat{h}(K)]^\mathrm{T}\in\mathbb{C}^K$ with $K $ the number of frequency bins used.

The input to the VAE-SSL and the supervised CNN is a temporally-ordered RTF-phase sequence. The $n\text{th}$ RTF-phase sequence is 
\begin{align}
 \mathbf{x}_n=\text{vec}(\text{phase}(\widehat{\mathbf{H}}_n))\in\mathbb{R}^{KP},   
\end{align}
with $\widehat{\mathbf{H}}_n=[\widehat{\mathbf{h}}_n\dots \widehat{\mathbf{h}}_{n+P-1}]\in\mathbb{C}^{K\times P}$,  $K=N_\mathrm{STFT}/2-1$, and $P$ the number of RTF-phase frames in the sequence. We use the wrapped RTF-phase, which is in the interval $[-\pi,\pi]$ radians.

\begin{figure}[t]
\centering
\includegraphics[width = 2.5in]{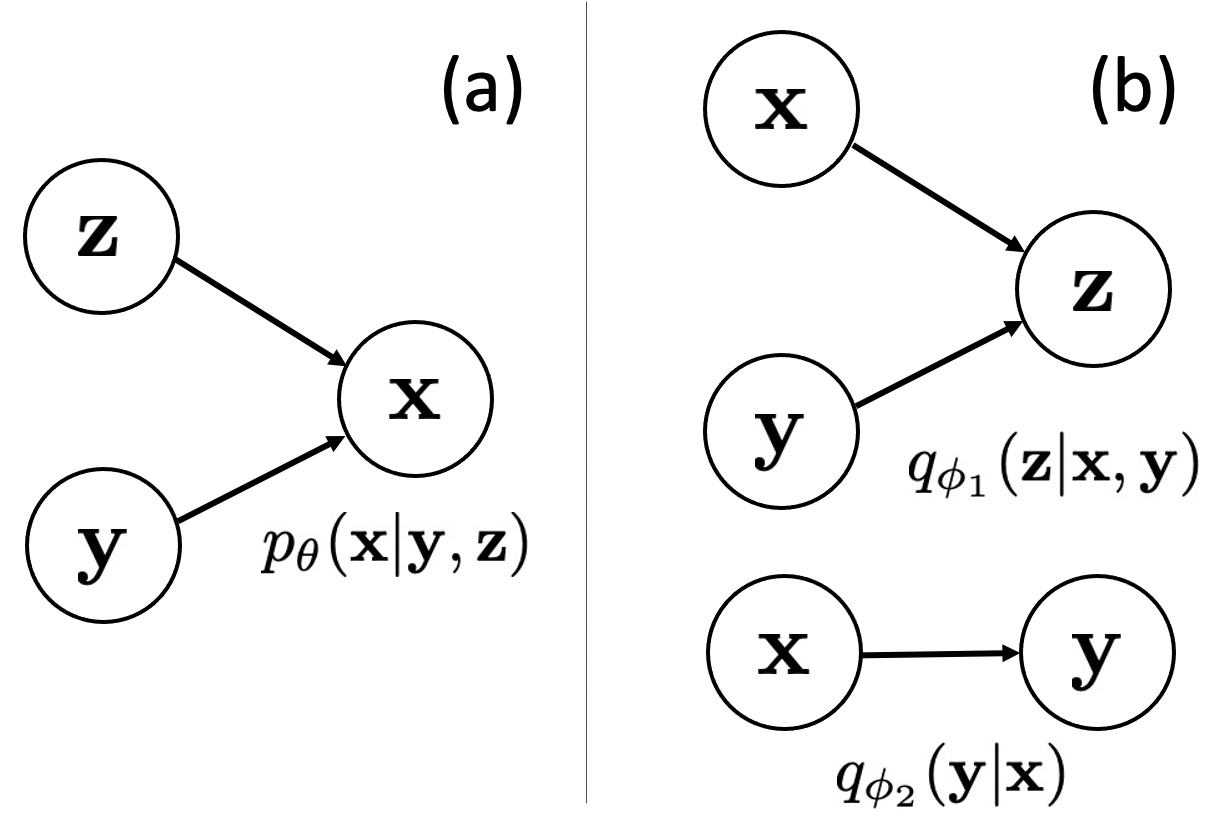}
\caption{Graphical models for VAE-SSL (a) generative and (b) inference models, with corresponding conditional distribution $p_\theta(\mathbf{x}|\mathbf{y,z})$.  In the generative model (a), the RTF sequence $\mathbf{x}$ is conditioned on the DOA $\mathbf{y}$ and the latent variable $\mathbf{z}$. We thus assume an inference model (b) to estimate them, in which $\mathbf{z}$ is conditioned on $\mathbf{x}$ and $\mathbf{y}$. Further, $\mathbf{y}$ is conditioned on $\mathbf{x}$, which represents the label inference (classifier) network. The generative model conditional distribution is parameterized by a decoder, see Fig.~\ref{fig:nn_config}(c) and the inference model conditional distributions parameterized by encoders, see Fig.~\ref{fig:nn_config}(a,b).}
\label{fig:ssvvae_dgm}
\end{figure} 

\subsection{Semi-supervised DOA estimation}
In acoustics we are often faced with scenarios where we have large volumes of acoustic recordings from arrays but potentially only few labels. The recordings themselves contain physical information, but the paucity of labels limit the task-specific value of this physical information. We here address this issue by formulating an SSL-based approach to source localization. We use a VAE model to obtain latent distributions of the RTF-phase physics for a particular room and corresponding DOAs from labeled and unlabeled examples. 

In our SSL formulation to DOA estimation, only a subset of the full dataset containing $N$ RTF sequences have corresponding DOA labels. Here the DOA labels are represented by $\mathbf{y}\in\{0,1\}^T$, a one-hot encoding, with $T$ the number of DOA classes. We thus have labeled and unlabeled sets defined by $\{\mathbf{x}_j,\mathbf{y}_j\}\in\mathcal{D}_l$ and $\{\mathbf{x}_{u}\}\in\mathcal{D}_u$. We further have labels for the unlabeled sequences $\mathbf{y}_{u}$, which are reserved to test the performance of the system only after training and validation. The sizes of the sets are $|\mathcal{D}_l|=J$ and $|\mathcal{D}_u|=N-J$. We thus have $J$ labeled RTF sequences and $N-J$ unlabeled sequences.

The $N-J$ unlabeled RTFs contain physical information, which is extracted by unsupervised learning. The $J$ labeled RTFs have both physical information and corresponding DOA labels, which help give the model task-specific value for DOA estimation by training the classifier. The generative model, which is trained on all samples helps guide the classifier training when labels are not available. Our goal is thus to well-infer the labels corresponding to labeled and unlabeled RTFs based on the trained VAE-SSL model.

\subsection{Semi-supervised learning  with VAEs}
We formulate a principled SSL framework based on VAEs\cite{kingma2014semi,kingma2019introduction}. A classifier NN and VAE are trained jointly, using both labeled and unlabeled data. The approach treats the label $\mathbf{y}$ as either latent and observed, depending on whether the data is labeled ($\{\mathbf{x}_j,\mathbf{y}_j\}\sim\mathcal{D}_l$) or unlabeled ($\{\mathbf{x}_u\}\sim\mathcal{D}_u$).  This corresponds to the `M2' model in \cite{kingma2014semi}. To simplify notation, we disregard subscripts for the theory derivation.

VAEs\cite{kingma2014semi,kingma2019introduction} combine the rich, high-dimensional modeling capacity of NNs with the tools of variational inference (VI)\cite{blei2017variational} to learn generative models of high-dimensional distributions. In VAEs, the distributions are parameterized by NNs, and the NN parameters are updated to best fit the assumed distributions to the data. We here adopt the VAE approach, and in the following the conditional probabilities are parameterized using NNs.

We assume each RTF phase sequence $\mathbf{x}$ is generated by a random process involving the latent random variable $\mathbf{z}\in\mathbb{R}^M$, with $M$ the dimension of the latent space, and source location label $\mathbf{y}$. We thus approximate the true data distribution $p^*(\mathbf{x})$ with the conditional distribution $p_\theta(\mathbf{x}|\mathbf{y},\mathbf{z})$, where $\theta$ are the parameters of the NN used to define the distribution. As will be discussed in Sec.~\ref{sec:parameterize}, we use $p_\theta(\mathbf{x}|\mathbf{y},\mathbf{z})=\overline{\mathcal{N}}(\cdot|\cdot)$, the truncated normal distribution with mean and variance defined by a NN with parameters (weights and biases) $\theta$. In the optimization stage, the parameters $\theta$ are adjusted to fit $p_\theta(\mathbf{x}|\mathbf{y},\mathbf{z})$ to the data.
% The mean and variance of this parametric distribution are $\boldsymbol{\mu}_\theta(\mathbf{y,z})$ and $\boldsymbol{\sigma}_\theta^2(\mathbf{y,z})$, and are the outputs of the decoder NN. 
$\mathbf{y}$ and $\mathbf{z}$ are assumed independent, with their marginal densities $p({\mathbf{y}})$ and $p(\mathbf{z})$ the categorical and normal distributions (see Sec.~\ref{sec:parameterize}). We thus have the generative model $p(\mathbf{x,y,z})=p_\theta(\mathbf{x|y,z})p(\mathbf{y})p(\mathbf{z})$, giving the graphical model Fig.~\ref{fig:ssvvae_dgm}(a).
% \begin{align}
% p_\theta({\mathbf{y}})=\mathrm{Cat}(\mathbf{y}|\boldsymbol{\pi});~p_\theta(\mathbf{z})=\mathcal{N}(\mathbf{0},\mathbf{I})
% \end{align}
% $\mathrm{Cat}(\cdot|\cdot)$ the categorical (multinomial) distribution with $\boldsymbol{\pi}\in\mathbb{R}^T$ the probabilities of the classes (which are assumed equal    
% p_\theta({\mathbf{y}})=\mathrm{Cat}(\mathbf{y}|\boldsymbol{\pi})$, with $\boldsymbol{\pi}\in\mathbb{R}^T$ the probabilities of the classes (which are assumed equal and normalized such that $\sum_T\pi_t=1$); and  $p_\theta(\mathbf{z})=\mathcal{N}(\mathbf{0},\mathbf{I})$.

Now we are presented with the challenge of inferring $\mathbf{y}$ (when it is not specified) and the latent variable $\mathbf{z}$. For labeled data $\{\mathbf{x}_j,\mathbf{y}_j\}\sim\mathcal{D}_l$,  the posterior of the latent variable $\mathbf{z}$ is
\begin{align}\label{eq:bayes2}
p(\mathbf{z}|\mathbf{x},\mathbf{y})=\frac{p(\mathbf{x},\mathbf{y}|\mathbf{z})p(\mathbf{z})}{p(\mathbf{x,y})}
\end{align}
and for unlabeled data $\{\mathbf{x}_u\}\sim\mathcal{D}_u$, the joint posterior of the latent  $\mathbf{z}$ and the label $\mathbf{y}$ is
\begin{align}\label{eq:bayes3}
p(\mathbf{y},\mathbf{z}|\mathbf{x})=\frac{p_\theta(\mathbf{x}|\mathbf{y},\mathbf{z})p(\mathbf{y},\mathbf{z})}{p(\mathbf{x})}.
\end{align}

\begin{figure*}[ht]
\centering
\includegraphics[width = 5in]{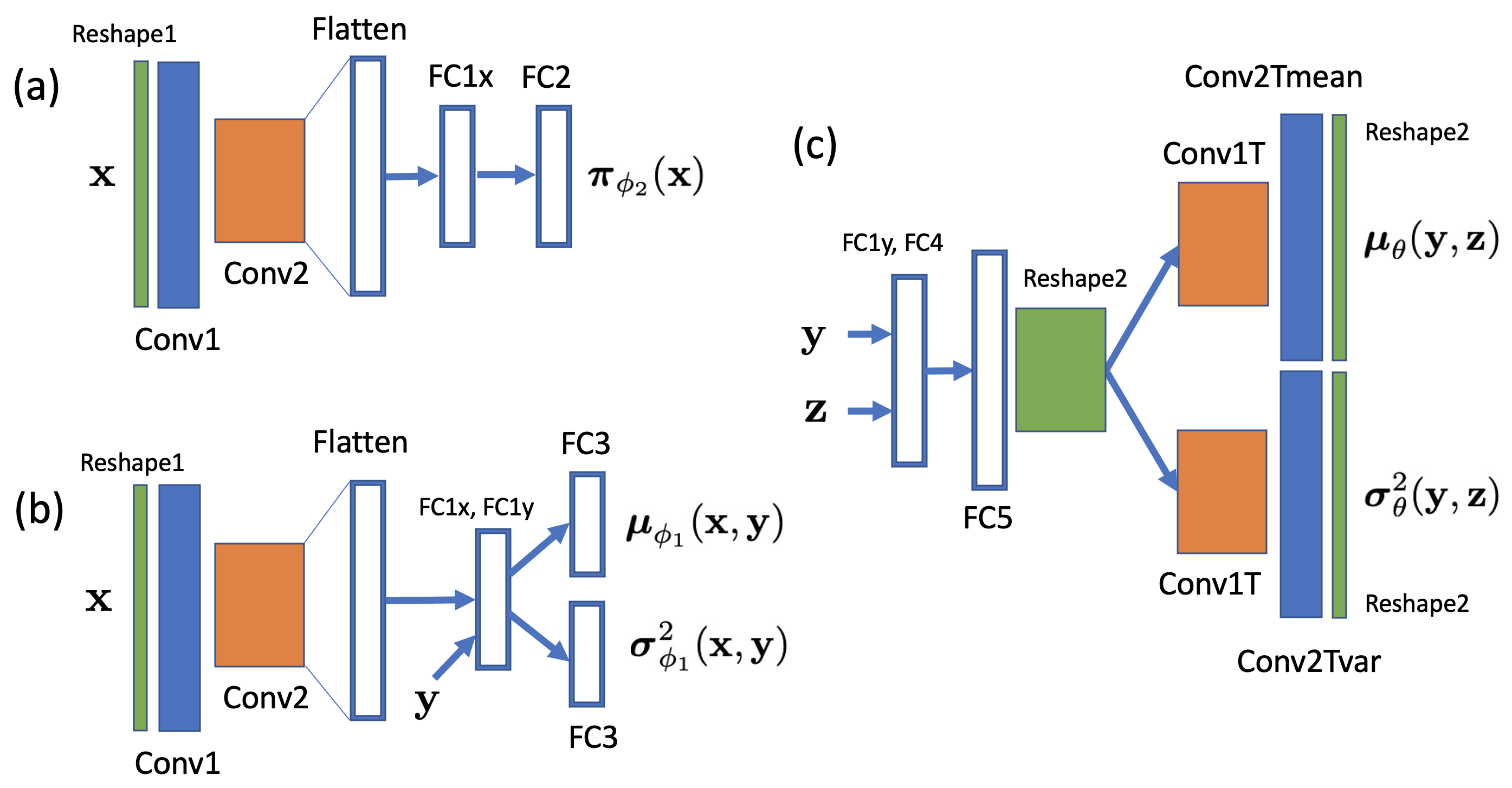}
\caption{Neural network configurations. Encoders for (a) label inference (classifier) and (b) latent model. (c) Decoder for generative model.}
\label{fig:nn_config}
\end{figure*}

Direct estimation of the posterior, e.g.\ from \eqref{eq:bayes2}, $p(\mathbf{z}|\mathbf{x},\mathbf{y})$ is nearly always intractable due to $p(\mathbf{x},\mathbf{y})=\int p(\mathbf{x},\mathbf{y},\mathbf{z})d\mathbf{z}$. Thus the posteriors are approximated using VI. We define a variational distribution which approximates the intractable posterior $p(\mathbf{z}|\mathbf{x},\mathbf{y})$, $q_{\phi_1}(\mathbf{z}|\mathbf{x},\mathbf{y})\approx p(\mathbf{z}|\mathbf{x},\mathbf{y})$. The variational distribution $q_{\phi_1}(\mathbf{z}|\mathbf{x},\mathbf{y})$ is a family of distributions parameterized by the latent inference encoder, with parameters $\phi_1$ (see Fig.~\ref{fig:nn_config}(b)). For $q_{\phi_1}(\mathbf{z}|\mathbf{x},\mathbf{y})$, we use a normal distribution. The graphical model for the corresponding inference model is shown in Fig.~\ref{fig:ssvvae_dgm}(b). The inference model for the DOA label $q_{\phi_2}(\mathbf{y|x})$, parameterized by label inference encoder with parameters $\phi_2$ (see Fig.~\ref{fig:nn_config}(a), \eqref{eq:label_k4} and \eqref{eq:label_k5}) is obtained starting in \eqref{eq:label_k4}. For $q_{\phi_2}(\mathbf{y|x})$, we use the categorical distribution. Similar to the generative model, in the optimization stage, the inference model NN parameters $\Phi=\{\phi_1,\phi_2\}$ are adjusted so the assumed distributions best model the data. The distributions are defined in Sec.~\ref{sec:parameterize}.

Starting with the model for the labeled data, see \eqref{eq:bayes2}, per VI we seek $q_\phi(\mathbf{z|x,y})$ which minimizes the KL-divergence
\begin{align}\label{eq:kl2}
\{\phi_1,\theta\}=\underset{\phi_1,\theta}{\arg\min}~\mathrm{KL}(q_{\phi_1}(\mathbf{z|x,y})||p(\mathbf{z|x,y})).
\end{align}
Considering first the labeled data $\mathcal{D}_l$, assessing the $\mathrm{KL}$-divergence, we obtain from \eqref{eq:kl2}
\begin{align}\label{eq:label_kl}
\mathrm{KL}&(q_{\phi_1}(\mathbf{z}|\mathbf{x},\mathbf{y})||p(\mathbf{z}|\mathbf{x},\mathbf{y})) \nonumber \\
&=\mathbb{E}\big[\log q_{\phi_1}(\mathbf{z}|\mathbf{x},\mathbf{y})\big]-\mathbb{E}\big[\log p(\mathbf{x},\mathbf{y}|\mathbf{z})p(\mathbf{z})\big] \\ \nonumber
&\hspace{5ex} + \log p(\mathbf{x,y}) \\ \nonumber
&=-\mathrm{ELBO}+ \log p(\mathbf{x,y}) \ge 0,
\end{align}
with the expectation $\mathbb{E}$ relative to $q_{\phi_1}(\mathbf{z}|\mathbf{x},\mathbf{y})$. This reveals the dependence of the KL divergence on evidence $p(\mathbf{x,y})$, which is intractable. The other two terms in \eqref{eq:label_kl} form the evidence lower bound (ELBO). Since the KL divergence is non-negative, the ELBO `lower bounds' the evidence: $\text{ELBO}\le\log p(\mathbf{x,y})$.
% Maximizing the ELBO is equivalent to minimizing the $\mathrm{KL}$ \eqref{eq:kl2}. 
We thus minimize $-\text{ELBO}$.

Considering the ELBO terms from \eqref{eq:label_kl}, we formulate the objective for labeled data, with $\mathbf{y}$ and $\mathbf{z}$ independent (in the generative model, see Fig.~\ref{fig:ssvvae_dgm}(a)) 
\begin{gather}
\begin{aligned}
-C(\theta,\phi_1;\mathbf{x,y})&=\mathbb{E}\big[\log p(\mathbf{x},\mathbf{y}|\mathbf{z})p(\mathbf{z})-\log q_{\phi_1}(\mathbf{z}|\mathbf{x},\mathbf{y})\big] \\
&=\mathbb{E}\big[\log p_\theta(\mathbf{x}|\mathbf{y},\mathbf{z}) + \log p(\mathbf{y})+ \log p(\mathbf{z}) \\
& \hspace{5ex} -\log q_{\phi_1}(\mathbf{z}|\mathbf{x},\mathbf{y})\big].
\end{aligned} \raisetag{0.8\baselineskip}
\label{eq:label_k2}
\end{gather}
This follows [Ref.~\cite{kingma2014semi}, Eq.(6)].

Next, an objective for unlabeled data $\mathcal{D}_u$ is derived. The intractable posterior from \eqref{eq:bayes3} is approximated by $q_\Phi(\mathbf{y},\mathbf{z}|\mathbf{x})\approx p(\mathbf{y},\mathbf{z}|\mathbf{x})$. From the KL-divergence, we find the objective (negative ELBO), using terms from \eqref{eq:bayes3}, as
\begin{gather}
\begin{aligned}
-&D(\theta,\Phi;\mathbf{x})= \mathbb{E}_{q_{\phi_2}(\mathbf{y}|\mathbf{x})}\big[ \mathbb{E}_{q_{\phi_1}(\mathbf{z}|\mathbf{x},\mathbf{y})} \big[ \log p_\theta(\mathbf{x}|\mathbf{y},\mathbf{z}) \\ 
& \hspace{10ex} +\log p(\mathbf{y})+\log p(\mathbf{z})-\log q_{\phi_1}(\mathbf{z}|\mathbf{x},\mathbf{y}) \\
&\hspace{10ex}  -\log q_{\phi_2}(\mathbf{y}|\mathbf{x})\big]\big]\\
& =\sum_\mathbf{y} q_{\phi_2}(\mathbf{y}|\mathbf{x})\big[-C(\theta,\phi;\mathbf{x,y})-\log q_{\phi_2}(\mathbf{y}|\mathbf{x})\big]
\end{aligned} \raisetag{1.4\baselineskip}
\label{eq:label_k4}
\end{gather}
This follows [Ref.~\cite{kingma2014semi}, Eq.(7)]. More details of the derivation are given in Appendix~\ref{app:unsup}. 

Assessing the terms in the supervised learning objective $C(\theta,\phi_1;\mathbf{x,y})$ \eqref{eq:label_k2}, it does not condition the label $\mathbf{y}$ on the sample $\mathbf{x}$, since for the supervised case the label is assumed known per \eqref{eq:bayes2}. The term $\log q_{\phi_2}(\mathbf{y}|\mathbf{x})$ is only present in the unsupervised learning objective \eqref{eq:label_k4}.  In this configuration, the classifier network learns only from the unsupervised sequence. It is important for the classifier to learn from the labeled sequences and we enforce this by adding an auxiliary term $-\log q_{\phi_2}(\mathbf{y}|\mathbf{x})$ to the supervised objective. This is a typical procedure \cite{kingma2014semi}.

An overall objective for training the VAE and classifier models using labeled and unlabeled data is derived by combining \eqref{eq:label_k2} and \eqref{eq:label_k4} with the auxiliary term
\begin{align}\label{eq:label_k5}
\mathcal{L} = \sum_{\{\mathbf{x}_j,\mathbf{y}_j\}\sim\mathcal{D}_l} &C(\theta,\phi_1;\mathbf{x}_j,\mathbf{y}_j) - \alpha \log q_{\phi_2}(\mathbf{y}_j|\mathbf{x}_j) \nonumber \\ 
&+ \sum_{\{\mathbf{x}_u\}\sim\mathcal{D}_u} D(\theta,\Phi;\mathbf{x}_u),
\end{align}
%
% %
% \begin{align}\label{eq:label_k6}
% \mathcal{J}^\alpha = \mathcal{J} + \alpha \sum_{\{\mathbf{x,y}\}} -\log q_\phi(\mathbf{y}|\mathbf{x}).
% \end{align}
% %
with $\alpha$ a scaling term, selected by hyperparameter optimization. This follows [Ref.~\cite{kingma2014semi}, Eqs.(8,9)]. 

\subsection{Distributions and neural network (NN) parameters}
\label{sec:parameterize}
The VAE-SSL model consists of 3 convolutional NNs which, for training and conditional generation, parameterize the probability distributions in \eqref{eq:label_k2} and \eqref{eq:label_k4}. The networks correspond to the inference model, with parameters $\Phi$, and the generative model, with parameters $\theta$. See Fig.~\ref{fig:nn_config} and Table.~\ref{table:nn_config} for NN configurations, and Sec.~\ref{sec:learning_param} for further details on implementation. 

For the inference model, we specify the following distributions, parameterized by NNs:
\begin{itemize}
    \item $q_{\phi_1}(\mathbf{z}|\mathbf{y,x})=\mathcal{N}(\mathbf{z}|\boldsymbol{\mu}_{\phi_1}(\mathbf{x,y}),\mathrm{diag}(\boldsymbol{\sigma}_{\phi_1}^2(\mathbf{x,y})))$, with $\mathcal{N}(\cdot|\cdot)$ the normal distribution parameterized by the outputs of the latent inference network $\boldsymbol{\mu}_{\phi_1}(\mathbf{x,y})\in\mathbb{R}^M$ and $\boldsymbol{\sigma}_{\phi_1}^2(\mathbf{x,y})\in\mathbb{R}^M$ (see Fig.~\ref{fig:nn_config}(b)).
    \item $q_{\phi_2}(\mathbf{y|x})=\mathrm{Cat}(\mathbf{y}|\boldsymbol{\pi}_{\phi_2}(\mathbf{x}))$, with $\mathrm{Cat}(\cdot|\cdot)$ the categorical (multinomial) distribution parameterized by the output of the classifier network $\boldsymbol{\pi}_{\phi_2}(\mathbf{x})\in\mathbb{R}^T$ (see Fig.~\ref{fig:nn_config}(a)).
\end{itemize}

For the generative model, we specify the distribution as
\begin{itemize}
    \item $p_\theta(\mathbf{x}|\mathbf{y},\mathbf{z})=\overline{\mathcal{N}}(\mathbf{x}|\boldsymbol{\mu}_\theta(\mathbf{y,z}),\mathrm{diag}(\boldsymbol{\sigma}_\theta^2(\mathbf{y,z})))$, with $\overline{\mathcal{N}}(\cdot|\cdot)$ the truncated normal distribution parameterized by the outputs of the decoder $\boldsymbol{\mu}_\theta(\mathbf{y,z})\in\mathbb{R}^{KP}$ and $\boldsymbol{\sigma}_\theta^2(\mathbf{y,z})\in\mathbb{R}^{KP}$ (see Fig.~\ref{fig:nn_config}(c)).
\end{itemize}

% 1 ) the label inference (classifier) network $\boldsymbol{\pi}_\phi(\mathbf{x})\in\mathbb{R}^T$, the output of which is used to parameterize $q_\phi(\mathbf{y|x})$; 2) the latent inference network $\boldsymbol{\mu}_\phi(\mathbf{x,y})\in\mathbb{R}^M$ and $\boldsymbol{\sigma}_\phi^2(\mathbf{x,y})\in\mathbb{R}^M$, parameterizing $q_\phi(\mathbf{z}|\mathbf{y,x})$; and 3) the decoder (generative) network $\boldsymbol{\mu}_\theta(\mathbf{y,z})\in\mathbb{R}^{KP}$ and $\boldsymbol{\sigma}_\theta^2(\mathbf{y,z}))\in\mathbb{R}^{KP}$ which parameterizes $p_\theta(\mathbf{x}|\mathbf{y},\mathbf{z})$.

% The densities parameterized by the NNs are for label inference  $q_\phi(\mathbf{y|x})=\mathrm{Cat}(\mathbf{y}|\boldsymbol{\pi}_\phi(\mathbf{x}))$ with $\mathrm{Cat}(\cdot|\cdot)$ the categorical (multinomial) distribution and for latent inference the normal distribution $q_\phi(\mathbf{z}|\mathbf{y,x})=\mathcal{N}(\mathbf{z}|\boldsymbol{\mu}_\phi(\mathbf{x,y}),\mathrm{diag}(\boldsymbol{\sigma}_\phi^2(\mathbf{x,y})))$. 

We use the truncated normal distribution for the generative conditional distribution $p_\theta(\mathbf{x}|\mathbf{y,z})$ since the wrapped RTF phase is on the interval $[-\pi,\pi]$. 

The marginal densities are defined as $p({\mathbf{y}})=\mathrm{Cat}(\mathbf{y}|\boldsymbol{\pi})$, with $\boldsymbol{\pi}\in\mathbb{R}^T$ the probabilities of the classes, which are assumed equal with $\pi_t=1/T$; and  $p(\mathbf{z})=\mathcal{N}(\mathbf{0},\mathbf{I})$.

Since the decoder parameterizes a truncated normal density, hyperbolic tangent activation is applied to the outputs of the decoder to constrain the mean, and a scaled sigmoid activation is applied to the decoder variance: $\boldsymbol{\mu}_\theta(\mathbf{y}_i,\mathbf{z}_j)\in[-1,1]$ and variance $\boldsymbol{\sigma}_\theta^2(\mathbf{y}_i,\mathbf{z}_j)\in[0,10]$. The input to VAE-SSL is normalized by $\pi$, thus $[-1,1]$ corresponds to the interval $[-\pi,\pi]$. The variance is limited to improve the training speed, since the truncated normal is implemented using rejection sampling.

The optimization of the VAE-SSL parameters $\{\theta,\Phi\}$ from the objective \eqref{eq:label_k5} is performed using stochastic variational inference (SVI) \cite{kingma2014auto}. The neural networks are implemented using Pytorch\cite{pytorch} and SVI is here implemented with the probabilistic programming package Pyro\cite{bingham2018pyro}, which evaluates the ELBO gradients with Monte Carlo sampling from the ELBO terms directly (instead of reparameterizations, e.g. the reparameterization trick in \cite{kingma2014semi}). 

During each training epoch, minibatches of RTF-phase sequences are drawn from the exclusively labeled $\{\mathbf{x}_j,\mathbf{y}_j\}\sim\mathcal{D}_l$ or unlabeled $\{\mathbf{x}_u\}\sim\mathcal{D}_u$ sets. In the case of $\mathcal{D}_l$, the terms corresponding to $\mathcal{D}_l$ in \eqref{eq:label_k5} are evaluated. Similarly for $\mathcal{D}_u$, only the unlabeled terms are evaluated. Labeled minibatches were used approximately once per $N/J$ unlabeled batches. For each data sample in the minibatch, the ELBO terms are sampled once. This has been found to be sufficient, provided the minibatch size is large enough \cite{kingma2014semi} (we use minibatch size of 256, see Sec.~\ref{sec:learning_param}). 

\begin{table*}[ht]
\caption{Network parameters corresponding to configuration in Fig.~\ref{fig:nn_config}. Since the kernel size is 3, and stride is 2 in the convolution layers, this gives transformation of input dimension $m$ as $w(m)=(m-1)/2$.}\label{table:nn_config}. 
  \centering
  \normalsize
  \begin{tabular}{l | l | l | l | l | l }
    \hline
    Name & Type & Input shape & Output shape & Kernel & Activation \\ 
    \hline\hline
    Reshape1  & Reshape                & $[K\times P]$      & $[K,P,1]$   & --     & --     \\
    Conv1     & Convolution            & $[K,P,1]$   & $[w(K),w(P),32]$               & $[3,3]$  & ReLU   \\
    Conv2     & Convolution            & $[w(K),w(P),32]$  & $[w(w(K)),w(w(P)),64]$   & $[3,3]$  & ReLU   \\
    Flatten   & Reshape                & $[w(w(K)),w(w(P)),64]$   & $[w(w(K))\times w(w(P))\times 64]$     & --     & --     \\
    FC1x      & Fully connected        & $[w(w(K))\times w(w(P))\times 64]$    & $[200]$       & --     & ReLU   \\
    FC1y      & Fully connected        & $[T] $        & $[200]$       & --     & ReLU   \\
    FC2       & Fully connected        & $[200]$       & $[T]$         & --     & Softmax   \\
    FC3       & Fully connected        & $[200] $      & $[50] $       & --     & None   \\
    FC4       & Fully connected        & $[50]$        & $[200] $      & --     & ReLU   \\
    FC5       & Fully connected        & $[200] $      & $[w(w(K))\times w(w(P))\times 64]$     & --     & ReLU   \\
    Reshape2  & Reshape                & $[w(w(K))\times w(w(P))\times 64]$ & $[w(w(K)), w(w(P)), 64]$   & --     & --   \\
    Conv1T    & Transpose conv.  & $[w(w(K)), w(w(P)), 64]$   & $[w(K), w(P), 32]$  & $[3,3]$  & ReLU  \\
    Conv2Tmean   & Transpose conv.  & $[w(K), w(P), 32] $ & $[K, P, 1]$  & $[3,3]$  & Tanh  \\
    Conv2Tvar   & Transpose conv.   & $[w(K), w(P), 32]$  & $[K, P, 1]$  & $[3,3]$  & Sigmoid  \\
    Reshape2  & Reshape                & $[K, P, 1]$   & $[K\times P] $   & --     & --     \\
    \hline
  \end{tabular}
\end{table*}

\subsection{Label estimation and conditional RTF generation}
\label{sec:label_gen}
From the trained inference model, the DOA is estimated by the indicator function
\begin{align}
y_{\rho n}=\mathbf{1}_{\rho=\widehat{t}},
\end{align}
with
\begin{align}
\widehat{t}=\underset{t}{\arg\max} (\pi_{t,{\phi_2}}(\mathbf{x}_n)),
\end{align}
% \begin{matrix}
% \ 1 &~\text{if}\ \ \widehat{t}=\underset{t}{\arg\max} (\pi_{t,\phi}(\mathbf{x}_n)) \\
% \ 0 &~\text{otherwise},
% \end{matrix}
% % \mathrm{max}(\boldsymbol{\pi}_\phi(\mathbf{x}_n)).
% \end{align}
%
% \begin{align}
% y_{\widehat{t}n}=\bigg\{
% \begin{matrix}
% \ 1 &~\text{if}\ \ \widehat{t}=\underset{t}{\arg\max} (\pi_{t,\phi}(\mathbf{x}_n)) \\
% \ 0 &~\text{otherwise},
% \end{matrix}
% % \mathrm{max}(\boldsymbol{\pi}_\phi(\mathbf{x}_n)).
% \end{align}
and $\rho$ and $t$ the discrete DOA indices, and $\pi_{t,\phi_2}(\mathbf{x}_n))$ the $t$-th output of $\boldsymbol{\pi}_{\phi_2}(\mathbf{x}_n))$. 
% The DOA angle represented by the one-hot coding $y_t$ is $A_t$. 
The DOA angle represented by the one-hot encoding $\mathbf{y}_n$ with active index $t$ is $D_{tn}$ and its estimate $\widehat{D}_{tn}$.

RTF-phase sequences can be conditionally generated from the trained generative model for a given label $\mathbf{y}_i$ using the prior $p(\mathbf{z}_j)$ and density $p_\theta(\mathbf{x}_j|\mathbf{y}_i,\mathbf{z}_j)$:
\begin{align} \label{eq:cond_sample}
    \mathbf{z}_j &\sim p(\mathbf{z}_j) \nonumber \\
    \mathbf{x}_j &\sim p_\theta(\mathbf{x}_j|\mathbf{y}_i,\mathbf{z}_j).
\end{align}

RTF-phase sequences can also be reconstructed (approximated) using the trained inference and generative models. For a given RTF-phase sequence $\mathbf{x}_i$ and label $\mathbf{y}_j$, $\mathbf{z}_j$ is sampled from the inference model
\begin{align} \label{eq:cond_sample2}
    \mathbf{z}_j &\sim q_{\phi_1}(\mathbf{z}_j|\mathbf{x}_j,\mathbf{y}_j),
\end{align}
then $\widehat{\mathbf{x}}_i$ is sampled from the generative model using $y_j$ and $z_j$
\begin{align} \label{eq:cond_sample3}
    \widehat{\mathbf{x}}_j &\sim p_\theta(\mathbf{x}_j|\mathbf{y}_j,\mathbf{z}_j).
\end{align}

\section{Experiments}
\label{sec:exper}

We assess the DOA estimation performance of the VAE-SSL approaches in moderately reverberant environments against three alternative techniques: SRP-PHAT\cite{dibiase2001robust}, MUSIC\cite{schmidt1986multiple}, and a supervised CNN baseline. The performance of the methods, summarized in Table~\ref{table:results} and \ref{table:off_results}, is quantified in terms of mean absolute error (MAE) and sequence-level accuracy (Acc.). We further-assess the generative modeling capabilities of the trained VAE-SSL system, as shown in Figs.~\ref{fig:rtf_gen1}--\ref{fig:ifft_big}.

In the following, we define the MAE as
\begin{align}
\text{MAE}=\frac{1}{N-J}\sum_{\mathcal{D}_u}~|D_{tu}-\widehat{D}_{tu}|,
\end{align}
with $|\cdot|$ denoting the absolute value. The sequence level accuracy (Acc.) is defined by
\begin{align}
\text{Acc.}=\frac{100}{N-J}\sum_{\mathcal{D}_u}~\mathbf{1}_{D_{tu}=\widehat{D}_{tu}}.
\end{align}
% %
% with
% \begin{align}
% F(A_{tu},\widehat{A}_{tu})=\bigg\{
% \begin{matrix}
% \ 1 &~\text{if}\ \ \widehat{A}_{tu} = A_{tu}\\
% \ 0 &~\text{otherwise}.
% \end{matrix}
% \end{align}

We consider training and validation of the VAE-SSL approach using speech. This is to obtain real-world application performance. We use measured IRs experimentally obtained at the Technical University of Denmark (DTU). We also compare performance with the Multichannel Impulse Response (MIR) Database\cite{hadad2014multichannel}.

% We analyze the generated RTF from the VAE to help qualify the physics learned by the generative model, shown in Fig.~\ref{fig:reverb_exp}. 

\begin{figure}[t]
\hspace{-1ex}
\centering
\includegraphics[width = 3.2in]{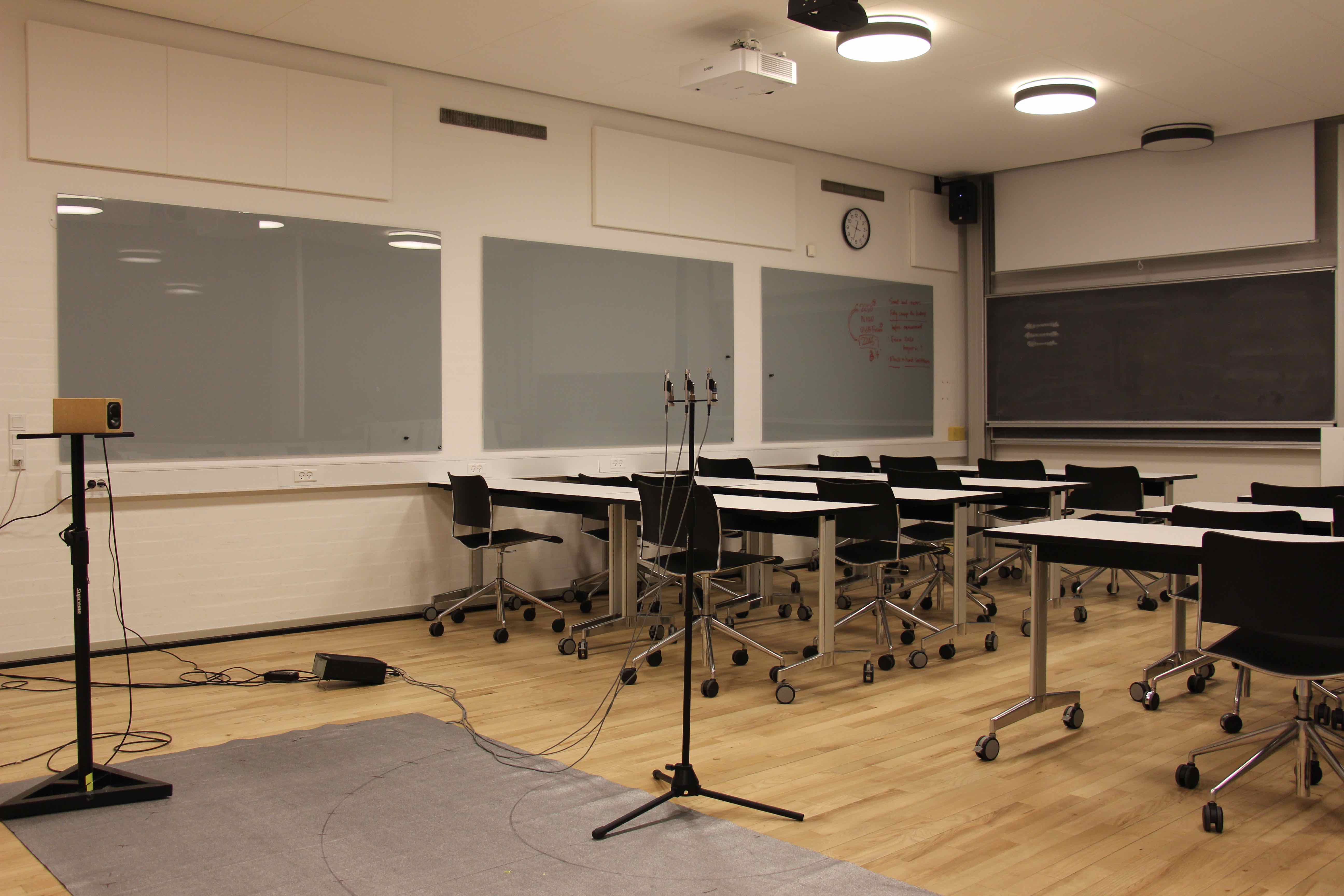}
\caption{Classroom and measurement configuration at Technical University of Denmark where room IRs were collected.}
\label{fig:classroom}
\end{figure}

\subsection{Measured Impulse Responses}
\label{sec:measurements}
We use measured IRs from two different datasets: a dataset recently recorded at DTU and the MIR database\cite{hadad2014multichannel}. We here briefly describe the experimental configuration for the DTU dataset. The data were recorded in a classroom at DTU in June 2020 (Fig.~\ref{fig:classroom}). The classroom was approximately rectangular, of dimensions $9\times 6 \times 3~\text{m}$ and fully furnished. One of the sidewalls is irregular, with a $100\times 40$~cm extrusion (for heating and ventilation) across the entire wall, in addition to support columns. All walls have scattering elements mounted on them (whiteboards, blackboards, diffusers and windowpanes), and the sound reflections are not specular. The nominal source-array range was 1.5 m. IRs were obtained from 19 DOAs ($10^\circ{}$ resolution over the interval [$-90^\circ{},90^\circ{}$]).  The reverberation time in the classroom was estimated to be $\text{RT}_{60}=500~\text{ms}$. There were two microphones, with 8.5~cm spacing. The sampling rate was 48~kHz. The IRs were truncated to 1~s and downsampled to 16~kHz for this study.

In addition to the nominal source grid for the DTU dataset, several off-range IRs (3 cases) and off-grid IRs (6 cases) were obtained to test the generalization of learning-based localization methods. The off-grid source DOAs were $[25^\circ{},28^\circ{},45^\circ{}]$ with $1.5$~m range. The off-range source DOAs (ranges) were $0^\circ{}$ ($1$~m), $10^\circ{}$ ($2$~m), $40^\circ{}$ ($2$~m), $-30^\circ{}$ ($2.5$~m), $-40^\circ{}$ ($2.5$~m), and $-30^\circ{}$ ($3.0$~m).

The MIR database was recorded in a $6\times6\times2.4$~m room with reverberation time controlled by acoustic panels \cite{hadad2014multichannel}. The dataset consists of source DOAs on a $15^\circ{}$ resolution over [$-90^\circ{},90^\circ{}$], giving 13 DOAs. Each of the DOAs was obtained at two ranges (1 and 2~m) with three reverberation times ($\text{RT}_{60}=160,360,~\text{and}~610$~ms). The IRs were obtained for 8 microphones located in the center of the room. There were several configurations, with different microphone spacing. We used data with 3 and 8~cm spacing. The sampling rate for the MIR database was also 48~kHz. For this study we use the two center microphones, with $8$~cm spacing. Further, from MIR we only use the $2$~m source range, with reverberation times $\text{RT}_{60}=360~\text{and}~610$~ms. The MIR IRs were processed in the same manner as the DTU IRs. For more details of the MIR database, see \cite{hadad2014multichannel}.

\subsection{RTF calculations}
The signal at the microphones is given in \eqref{eq:mics}. 
% In this study, for the DTU dataset, we use the two microphones located at (-8.5,0)~cm, (0,0)~cm. For the MIR dataset, we used the two center microphones, with $8$~cm spacing. Further, from MIR we only use the $2~m$ source range, with reverberation times $\text{RT}_{60}=360~\text{and}~610$~ms.
We obtain the RTFs from the data by \eqref{eq:mics_rtf_fft2}. The RTFs are estimated using single STFT frames with Hamming windowing with 50\% overlap and segment length $N_{FFT} = 256$. The VAE and the supervised CNN inputs $\mathbf{x}_n$ use $P=31$ RTF vectors, giving an input size $K\times P= 127\times 31$ (neglecting the highest frequencies from full RTF with length $N_\mathrm{FFT}/2+1$, including all frequencies between 0 (DC) up to $N_\mathrm{FFT}/2$, to support strided transpose convolution without padding). For fair comparison, SRP-PHAT and MUSIC used the $P=31$ STFT frames to estimate the RTFs, with $K=N_\mathrm{FFT}/2+1$. Thus, the temporal length of the sequences for all methods was $0.26$~s. 

\subsection{Speech data processing} \label{sec:speech_data}
We used speech data from the LibriSpeech corpus \cite{panayotov2015librispeech} development set, which contains 5.4 hours of recorded speech with a 16~kHz sampling rate from public domain audiobooks. The dataset contains equal numbers of male and female speakers (20 each). 

The speech segments from LibriSpeech were convolved with the recorded room IRs to obtain reverberant speech. Voice activity detection (VAD) was performed on the dry speech before convolution with the IRs. We used the WebRTC project VAD system \cite{webrtc}, a popular open source VAD system based on pretrained Gaussian mixture models. The VAD settings used were 3 (most aggressive) with a 10~ms analysis window.

%%%%%%%%%%%%%%%%%%%%%%%%%%%%%%%
%% BIG RESULTS TABLE %%%%%%%%%%%%%%%%%%%
%%%%%%%%%%%%%%%%%%%%%%%%%%%%%%%

\begin{table*}
\centering
\caption{Localization performance of VAE-SSL, fully supervised CNN, SRP-PHAT and MUSIC on (a, b) DTU dataset and (c, d) MIR database. Training and validation DOA estimation performance given for unlabeled data $\mathcal{D}_u$. Performance quantified in terms of mean absolute error (MAE) and sequence level accuracy (Acc.).} \label{table:results}
\normalsize
\subfloat[DTU training data]{
\begin{tabular}{c | l | l | l | l }
    \cline{2-5}
     \multicolumn{1}{c|}{} & \multicolumn{2}{c|}{VAE-SSL} & \multicolumn{2}{c}{CNN}\\ \hline
     $J$ (\# labels) & MAE & Acc. & MAE & Acc.\\ \hline\hline
    114  & 6.34  & 56.2  & 31.5  & 40.3 \\
    247  & 3.38  & 75.5  & 25.4  & 49.9 \\
    494	 & 1.85  & 84.9	 & 20.4	 & 60.1 \\
    988	 & 1.36	& 88.4	& 17.8	& 63.4 \\
    1995 & 1.29	& 89.3	& 15.0	& 70.1 \\
    3990 & 1.03	& 91.6	& 12.5	& 74.6 \\
    7999 & 0.73 & 94.0  & 10.0  & 81.2  \\
     \hline\hline
     \multicolumn{1}{c|}{SRP-PHAT} & \multicolumn{1}{l|}{4.72} & \multicolumn{1}{l|}{66.1} & \multicolumn{2}{c}{}\\ 
     \multicolumn{1}{c|}{MUSIC} & \multicolumn{1}{l|}{22.1} & \multicolumn{1}{l|}{25.5} & \multicolumn{2}{c}{}\\ 
    %  \multicolumn{1}{c|}{SBL} & \multicolumn{1}{l|}{11.5} & \multicolumn{1}{l|}{36.6} & \multicolumn{2}{c}{} \\
     \cline{1-3}
  \end{tabular}
}
\qquad
\subfloat[DTU validation data]{
\begin{tabular}{c | l | l | l | l }
    \cline{2-5}
     \multicolumn{1}{c|}{} & \multicolumn{2}{c|}{VAE-SSL} & \multicolumn{2}{c}{CNN}\\ \hline
     $J$ (\# labels) & MAE & Acc. & MAE & Acc.\\ \hline\hline
    114  & 7.81  & 52.0  & 33.1  & 37.8 \\
    247  & 5.05  & 68.7  & 27.6  & 46.5 \\
    494	 & 3.56  & 76.0	 & 23.2	 & 55.6 \\
    988	 & 2.99	& 79.5	& 20.3	& 59.1 \\
    1995 & 3.01	& 79.9	& 17.8	& 64.8 \\
    3990 & 2.92	& 81.4	& 16.2	& 68.0 \\
    7999 & 3.00 & 80.5  & 14.3  & 73.7  \\
     \hline\hline
     \multicolumn{1}{c|}{SRP-PHAT} & \multicolumn{1}{l|}{5.99} & \multicolumn{1}{l|}{60.9} & \multicolumn{2}{c}{}\\ 
     \multicolumn{1}{c|}{MUSIC} & \multicolumn{1}{l|}{22.2} & \multicolumn{1}{l|}{24.9} & \multicolumn{2}{c}{}\\
    %  \multicolumn{1}{c|}{SBL} & \multicolumn{1}{l|}{12.8} & \multicolumn{1}{l|}{32.5} & \multicolumn{2}{c}{} \\
     \cline{1-3}
\end{tabular}
} \\
\subfloat[MIR training data]{
\begin{tabular}{c | l | l | l | l }
    \cline{2-5}
     \multicolumn{1}{c|}{} & \multicolumn{2}{c|}{VAE-SSL} & \multicolumn{2}{c}{CNN}\\ \hline
     $J$ (\# labels) & MAE & Acc. & MAE & Acc.\\ \hline\hline
    117  & 9.65  & 82.0  & 17.0  & 51.1 \\
    247  & 1.89  & 92.4  & 9.32  & 67.8 \\
    494	 & 1.50  & 92.8	 & 6.11	 & 77.1 \\
    988	 & 1.41	 & 93.4  & 4.69	& 82.3 \\
    1989 & 1.37	 & 93.8	 & 4.04	& 84.2 \\
    3991 & 1.38	 & 93.9	 & 3.01	& 88.5 \\
    7995 & 1.12  & 95.1  & 2.09  & 91.9  \\
     \hline\hline
     \multicolumn{1}{c|}{SRP-PHAT} & \multicolumn{1}{l|}{6.67} & \multicolumn{1}{l|}{71.5} & \multicolumn{2}{c}{}\\
     \multicolumn{1}{c|}{MUSIC} & \multicolumn{1}{l|}{43.2} & \multicolumn{1}{l|}{21.7} & \multicolumn{2}{c}{}\\
     \cline{1-3}
  \end{tabular}
}
\qquad
\subfloat[MIR validation data]{
\begin{tabular}{c | l | l | l | l }
    \cline{2-5}
     \multicolumn{1}{c|}{} & \multicolumn{2}{c|}{VAE-SSL} & \multicolumn{2}{c}{CNN}\\ \hline
     $J$ (\# labels) & MAE & Acc. & MAE & Acc.\\ \hline\hline
    117  & 12.9  & 69.3  & 16.3  & 47.6 \\
    247  & 3.93  & 81.4  & 9.99  & 60.9 \\
    494	 & 3.46  & 82.1	 & 7.68	 & 67.9 \\
    988	 & 3.25	&  83.3	& 6.36	& 72.9 \\
    1989 & 3.32	&  83.5	& 5.71	& 74.4 \\
    3991 & 3.61	&  82.2	& 5.31	& 76.2 \\
    7995 & 3.11 &  84.4  & 4.96  & 77.1  \\
     \hline\hline
     \multicolumn{1}{c|}{SRP-PHAT} & \multicolumn{1}{l|}{5.34} & \multicolumn{1}{l|}{74.6} & \multicolumn{2}{c}{}\\ 
     \multicolumn{1}{c|}{MUSIC} & \multicolumn{1}{l|}{36.3} & \multicolumn{1}{l|}{29.4} & \multicolumn{2}{c}{}\\
     \cline{1-3}
  \end{tabular}
}
\end{table*}

VAD was applied to the entire LibriSpeech development corpus, and a total of 40 speech segments 2-3~s in duration were randomly selected for training and validation, 
% {\color{blue} (can increase the number of sequences if we think it would strengthen the paper)}. 
20 segments each. This yielded $\sim110,000$ RTF sequence for the nominal DTU IRs and $\sim76,000$ sequences for the MIR IRs ($\sim5800$ sequences per DOA for both datasets). This further yielded $\sim 17,500$ and $\sim 35,000$ sequences for the DTU off-grid and off-range measurements using the validation speech. Given one active speaker recording segment of time length $t$, with 50\% STFT overlap, the number of RTF-phase sequences in the segment $N_{\rm seg}$ is
\begin{align}
N_{\rm seg}\approx \frac{2Tf_st}{N_\mathrm{FFT}}-P.
\end{align}

For training and validation, $J$ labeled sequences were drawn uniformly from the concatenated reverberant speech sequences (for each DOA). The remaining sequences from the training set were used for unsupervised learning. Sensor noise with  20 dB SNR was added to the microphone signals (see \eqref{eq:mics}). We consider a range of values for $J$, for evaluating the effect of the number of labelled sequences on the performance of the learning-based approaches. SRP-PHAT and MUSIC were evaluated using the full dataset of $N$ sequences, for each scenario.

\subsection{Learning-based model parameters} \label{sec:learning_param}
The VAE-SSL model (classifier, inference, and generative networks) were implemented using strided CNNs, with stride of 2 pixels. The network architectures are given in Fig.~\ref{fig:nn_config}, and the corresponding parameters are given in Table~\ref{table:nn_config}. Each NN used two convolutional layers without pooling, and two fully-connected layers. The convolutional layers help process the high-dimensional RTF-sequence data by enabling parameter-sharing between neighboring pixels \cite{goodfellow2016deep}. Since the one-hot DOA $\mathbf{y}$ is relatively low-dimensional, it does not necessitate processing by convolutional layers. Thus it is input directly to the first fully-connected layer in the encoder model (Fig.~\ref{fig:nn_config}(b)) after the convolutional processing of $\mathbf{x}$.

Before being processed by the CNNs (Fig.~\ref{fig:nn_config}(a,b)), the generated RTF-phase sequence vectors $\widehat{\mathbf{x}}_n\in\mathbb{R}^{KP}$ are reshaped to a matrix with dimensions $P\times K$, so the convolutional filters account for correlations between time and frequency bins. The output of the CNNs (Fig.~\ref{fig:nn_config}(c)) is reshaped to a vector with length $KP$.

Several subnetworks reuse the same architecture, hence the names are reused. However, weights for each implementation are trained independently - i.e. no weight sharing is applied. The latent variable $\mathbf{z}$ dimension for all experiments was $M=50$, assuming that the hidden representation must account for temporal variation of the RTF-phase sequences, and residual variation of the RTF from speech. Dropout with probability of 0.5 is applied to the large fully connected layers FC1x and FC5 (Fig.~\ref{fig:nn_config}, Table~\ref{table:nn_config}). 

For the fully-supervised CNN approach, we used the classifier CNN architecture from the VAE-SSL model. The fully-supervised CNN was trained using only the labeled data $\mathcal{D}_l$, without integration with the generative model. Like VAE-SSL, the CNN was implemented using Pytorch. 

All NNs were optimized using Adam\cite{kingma2014adam}. For all cases, the learning rate was 5e-5 and the minibatch size was 256. The default momentum and decay values (0.9,0.999) were used from the Pytorch implementation of Adam. The VAE-SSL has the tuning parameter, the auxiliary multiplier $\alpha$, which weights the supervised objective (see \eqref{eq:label_k5}). It was found in the experiments that the performance of VAE-SSL was not very sensitive to $\alpha$, though the best value per validation accuracy was found by grid search over the interval [500,10000].

\subsection{Non-learning: SRP-PHAT and MUSIC configuration}
The same STFT frames used to calculate the RTF-phase sequences for VAE-SSL and CNN were used by the conventional SRP-PHAT\cite{dibiase2001robust,do2007real} and MUSIC\cite{schmidt1986multiple,evers2020locata} approaches. There were $P$ STFT frames in the VAE-SSL and CNN method input sequences $\mathbf{x}$. Thus the conventional approaches used 13 candidate DOAs over $[-90^\circ,90^\circ]$ for the MIR IRs and 19 for the DTU IRs. We used the SRP-PHAT and MUSIC implementations from the Pyroomacoustics toolbox \cite{scheibler2018pyroomacoustics}. For MUSIC, the number of sources was set to one.

\begin{figure}[t]
\hspace{-4ex}
\includegraphics[width = 3.8in]{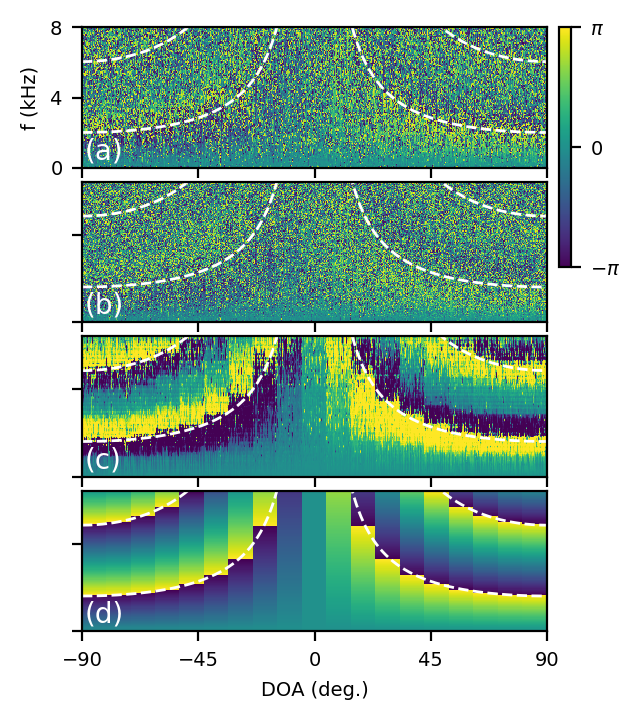}
\caption{Reconstruction of RTF phase sequences using VAE-SSL (trained using $J=988$ labels). 10 sequences are reconstructed for each DOA using the DTU IR data convolved with speech. (a) Input RTF phase sequences $\mathbf{x}_n$. (b) Reconstructed RTF phase sequences $\widehat{\mathbf{x}}_n$. (c) RTF phase sequence mean $\boldsymbol{\mu}_\theta(\mathbf{y}_i,\mathbf{z}_j)$. (d) Free space RTF per \eqref{eq:rtf_free}. Phase-wrap function plotted with white dashed lines. The color scale is RTF phase [$-\pi$,$\pi$] radians. Theory, other parameters located in Sec.~\ref{sec:theory}~D--E, Sec.~\ref{sec:exper}~D.}
\label{fig:rtf_gen1}
\end{figure}

\subsection{Training and localization performance}

For VAE-SSL and supervised CNN training, $J$ labeled sequences were drawn from the training set and $J$ labeled sequences were drawn from the validation set. The model was chosen based on the validation accuracy for labeled sequences. For VAE-SSL, the additional $N-J$ unlabeled sequences from the training dataset were used to train the networks with unsupervised learning. Only the labeled sequences were used to train the supervised CNN. After training, the performance of the models was evaluated using the unlabeled sequences from the validation dataset. Since for the unlabeled examples, the DOAs are technically unavailable, for unsupervised learning some RTF-phase sequences $\mathbf{x}_l$ contain frames with different DOA labels. The labeled sequences, and the unlabeled sequences used for performance evaluation, have only one DOA each. All the RTF-phase sequences $\mathbf{X}=[\mathbf{x}_1\dots\mathbf{x}_N]$ were normalized by $\pi$ for the VAE-SSL and CNN approaches.

\begin{table}[t]
\caption{Localization error (MAE) of VAE-SSL, fully supervised CNN, SRP-PHAT and MUSIC on off-grid and off-range measurements from DTU dataset.} \label{table:off_results}
\centering
\normalsize
\begin{tabular}{r | l | l }
\hline
Method & Off-grid & Off-range \\
\hline\hline
MUSIC    & 16.5       & 15.3 \\
SRP-PHAT & 6.29       & 4.16 \\
CNN	     & 8.96       & 47.7 \\
VAE-SSL	 & {\bf 6.15} &  {\bf 2.72} \\
RTF-1NN & 7.33 & 38.0 \\
\hline
\end{tabular}
\end{table}

The performance of VAE-SSL and the competing approaches for the DTU IR dataset are given in Table~\ref{table:results}(a,b). VAE-SSL and fully supervised CNN were trained using 20 speech signals and validated using an additional 20 speech signals (see Sec.~\ref{sec:speech_data}). We use $J=247,494,988,1995,3990,7999$ supervised sequences, which are multiples of the number of candidate DOAs ($T=19$) to ensure an equal number of labeled samples for each DOA. It is observed that the VAE-SSL approach generalizes to the validation data, with better performance than SRP-PHAT with as few as $J=247$ labeled sequences (or $13$ per DOA). For both the validation and training data, the performance increases significantly for additional labels. The  VAE-SSL outperforms CNN and MUSIC for all the experiments in this paper. It is apparent that in using the minimally processed RTF sequences, that our semi-supervised approach can learn to identify the relevant features from the data.

Similar trends are observed for the MIR database IRs, given in Table~\ref{table:results}(d,e). Here VAE-SSL and fully supervised CNN were trained and validated using the same 20 speech signals with two different reverberation times. The training data IR was $\textrm{RT}_{60}=0.610$~ms case, and the validation was $\textrm{RT}_{60}=0.360$~ms.  We use $J=247,494,988,1989,3991,7995$ supervised sequences, which are multiples of the number of candidate DOAs ($T=13$). It is observed that VAE-SSL significantly outperforms the fully supervised CNN, SRP-PHAT, and MUSIC methods for both the training and validation cases when the number of supervised examples is sufficient. A similar number of labels to the DTU dataset were required to outperform SRP-PHAT: $J=247$ labeled sequences, or in this case 19 labels per DOA. It is shown that the VAE-SSL generalizes well to different reverberation times.

We note that in the case of the DTU IR dataset, the CNN performs well in terms of accuracy (which overtakes SRP-PHAT), though the MAE remains large. For the MIR dataset, both CNN and MAE converge to that of SRP-PHAT. The acoustic environment for the DTU dataset, with classroom  furniture and irregular wall geometry, is more complex than the MIR database. This can be seen in the reduction of the accuracy for all approaches, despite the reduction in the number of DOAs (13 for DTU vs 19 for MIR). The CNN system cannot generalize well to the unseen data, since not enough labelled frames are available. While the accuracy is good, the mistakes by CNN are large. This is why we consider both metrics, since accuracy alone can be misleading. For VAE-SSL we find that the system generalizes well to the DTU environment using only few labels, since it extracts physically meaningful features as we will show in Sec.~\ref{sec:off} and \ref{sec:gen}.

\begin{figure}[t]
\hspace{-4ex}
\includegraphics[width = 3.8in]{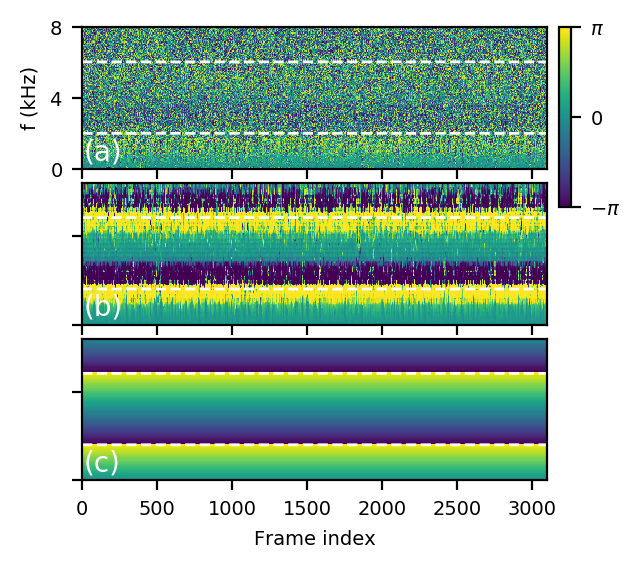}
\caption{Conditionally generated RTF phase sequences (100 sequences, 31 frames per sequence, 100*31=3100 frames) using VAE-SSL generative model trained with DTU IR data convolved with speech for DOA $90^\circ{}$ (trained using $J=988$ labels). (a) Conditionally generated RTF sequences $\widehat{\mathbf{x}}_j$. (c) Generated RTF phase sequence mean $\boldsymbol{\mu}_\theta(\mathbf{y}_i,\mathbf{z}_j)$. (d) Free space RTF per \eqref{eq:rtf_free}, same as Fig.~\ref{fig:rtf_gen1}.}
\label{fig:rtf_gen2}
\end{figure}

\begin{figure}[t]
\hspace{-4ex}
\includegraphics[width = 3.8in]{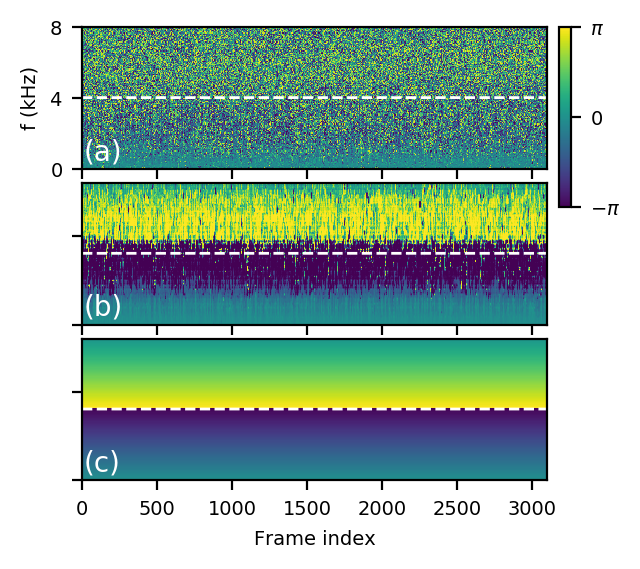}
\caption{Conditionally generated RTF phase sequences for DOA=$-30^\circ{}$ (trained using $J=988$ labels). Same configuration as Fig~\ref{fig:rtf_gen2}.}
\label{fig:rtf_gen3}
\end{figure}

\subsection{Off-grid and off-range generalization} \label{sec:off}
We previously quantified the performance of the localization methods for different speakers and reverberation times. We now assess the off-grid and off-range localization performance of the methods in a real-room environment with the DTU dataset. This important test examines whether the representations learned by the ML approaches, VAE-SSL and fully supervised CNN, can generalize to source locations that were not present in the training data. In real applications, it can be expected that the source locations will not precisely correspond to the training locations. The off-grid DOAs are $2$ to $5^\circ{}$ away from the training DOAs, and the off-range are $\pm 0.5,\text{and}~+1$ to $2$~m away, see Sec.~\ref{sec:exper}A.

For the learning-based approaches, results are given for systems trained with $J=988$ labels. For additional comparison, we consider a basic one-nearest-neighbors approach, which uses spectrally-averaged RTF features. This manual preprocessing should reduce the noise in the RTF features and give a reasonable baseline for a hand-crafted, but still data-driven, approach.

For the nearest-neighbors approach, which we deem RTF-1NN, we perform spectral averaging over the full RTFs \cite{laufer2016semi} using the STFT frames used to estimate the instantaneous RTF sequence (Sec.\ref{sec:rtfs}). The spectrally-averaged RTFs for each DOA are given by
\begin{align}
\widetilde{h}(k)=\frac{\sum_{s\in S} d_{2s}^{}d_{1s}^*}{\sum_{s\in S} d_{1s}^{}d_{1s}^*},
\end{align}
with $\cdot^{*}$ the complex conjugate and $S$ the set of STFT frames used. The RTF from the labeled data corresponding to each DOA is $\widetilde{h}_t(k)$, with $|S|=PJ/T$. The RTF from each sequence in the unlabeled data, which is to be classified, is $\widetilde{h}_u(k)$ with $|S|=P$. The DOA for $\widetilde{h}_u(k)$ is estimated by 
\begin{align}
    \underset{t}{\arg\min} \ \ \|\widetilde{\mathbf{h}}_t - \widetilde{\mathbf{h}}_u\|_2,
\end{align}
with $\|\cdot\|_2$ the $\ell_2$ norm. 
% We note that this formulation is related to matched-filter beamforming as suggested by \cite{jan1996sound}.

The results of the different methods are shown in Table~\ref{table:off_results}. We find that the VAE-SSL approach generalizes well to off-grid and off-range sources. It is shown that the trained VAE-SSL outperforms the other approaches. Again, this performance is achieved using minimal preprocessing of the input features. Particularly, for the off-range scenario, the CNN and RTF-1NN generalize quite poorly.

\subsection{RTF generation with VAE-SSL} \label{sec:gen}
The trained generative model from VAE-SSL can be used to generate new RTF sequences. We demonstrate RTF phase generation and discuss their physical interpretation. Generated RTF-phase sequences are shown in Fig.\ref{fig:rtf_gen1}--\ref{fig:rtf_gen3}. For display, the generated RTF-phase sequence vectors $\widehat{\mathbf{x}}\in\mathbb{R}^{KP}$ are reshaped to $P\times K$. We use the phase-wrap of the RTF ($\pm\pi$~rad.) (as a function of sensor separation and DOA $\theta_n$, $f=c/(2r|\sin\theta_n|)$) to help qualify the physics learned by VAE-SSL, with $r$ the microphone spacing. This is plotted along with the RTF-phase frames from the design case room configuration. We also compare the generated RTF-phases with the corresponding free-space RTF, which is calculated by
\begin{align} \label{eq:rtf_free}
\mathrm{phase}(H(f_k,\theta)) = \frac{2\pi f_k r}{c}\sin{\theta}.
\end{align}

\begin{figure}[t]
\hspace{-2ex}
\includegraphics[width = 3.5in]{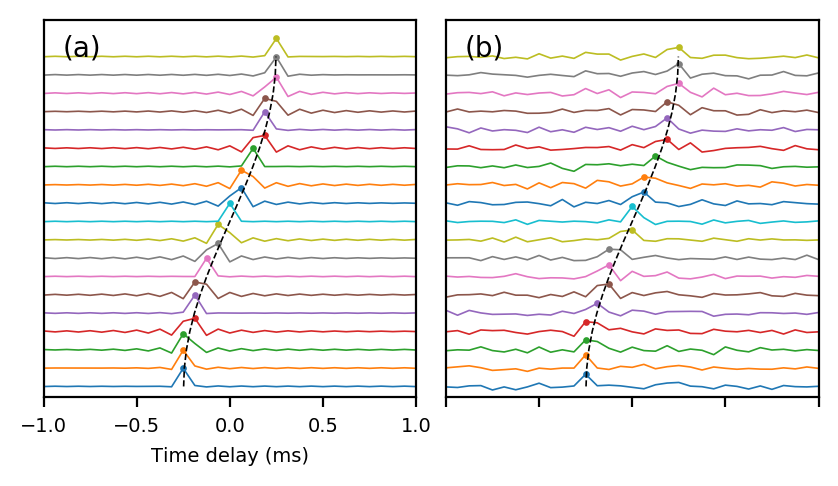}
\caption{Inverse fast Fourier transforms (IFFT) of RTF (from RTF phase), see Sec.~\ref{sec:phase_delay}. (a) Simulated anechoic time delay based on DTU dataset source-receiver parameters. (b) Time delays from reconstructed RTF-phase  mean from reverberant DTU dataset (see Fig.~\ref{fig:rtf_gen1}(b) for RTF-phase), calculated from one RTF realization per DOA. Hypothesized time delays, corresponding to the 19 DOAs in the dataset, are shown as black dashed line.}
\label{fig:rtf_ifft1}
\end{figure}

We use the VAE-SSL model trained on the DTU IRs to reconstruct and conditionally generate RTFs. In Fig.~\ref{fig:rtf_gen1}, 10 RTF sequences for each DOA ($T=19$ DOA, giving 190 sequences) are reconstructed. It is observed that the RTF phase is well-reconstructed, and that the reconstructed RTF phase (Fig.~\ref{fig:rtf_gen1}(b)) and phase mean (Fig.~\ref{fig:rtf_gen1}(c)) conform to the phase-wrap function. We further show the free-space RTF phase (per \eqref{eq:rtf_free}) in Fig.~\ref{fig:rtf_gen1}(d). The generated mean RTF phase correlates well with the free-space RTF phase.

Using the trained generative model from VAE-SSL with DTU IRs and speech, we conditionally sample (see Sec.~\ref{sec:label_gen}) for fixed label $\mathbf{y}$ and randomly drawn $\mathbf{z}_i$. In Fig.~\ref{fig:rtf_gen2}, we use $\mathbf{y}$ corresponding to DOA $90^\circ{}$ and sample 100 RTF phase frames. Similarly, in Fig.~\ref{fig:rtf_gen3} we use $\mathbf{y}$ corresponding to DOA $-30^\circ{}$. It is observed that the generated phase and its corresponding mean in each case is well-correlated with the predicted RTF phase wrap function. In the case of the conditionally generated RTFs, they also correlate well with their corresponding free-space RTFs. The output sampled from the generative model (see \eqref{eq:cond_sample}) is multiplied by $\pi$, since normalization was applied during training.

\begin{figure}[t]
\hspace{-2ex}
\includegraphics[width = 3.5in]{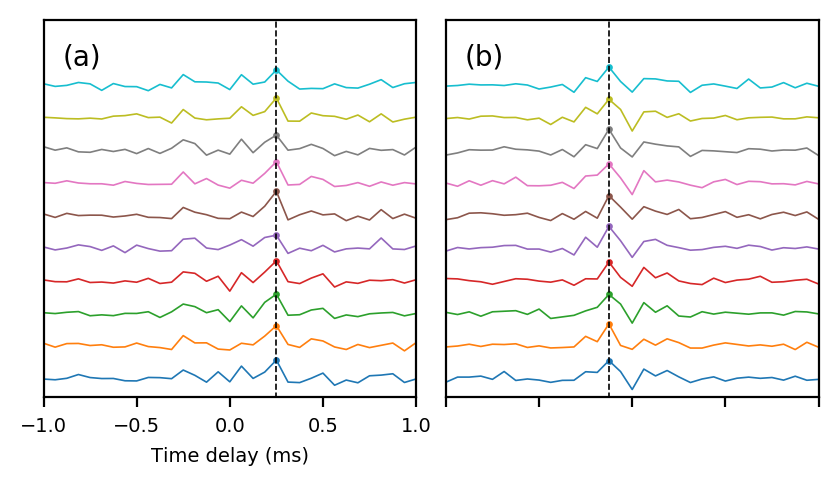}
\caption{IFFTs of VAE-SSL conditionally generated RTFs (time delay), see Sec.~\ref{sec:phase_delay}. Time delays from conditionally generated RTF-phase means from reverberant DTU dataset for (a) DOA = $90^\circ{}$ (see Fig.~\ref{fig:rtf_gen2}(b) for RTF-phase) and (b) DOA = $-30^\circ{}$ (see Fig.~\ref{fig:rtf_gen3}(b) for RTF-phase). Calculated for 10 RTF realizations (sampling random $\mathbf{z}$). Hypothesized time delay shown as black dashed line.}
\label{fig:rtf_ifft2}
\end{figure}

\subsubsection{RTF time delay}\label{sec:phase_delay}
We further evaluate the generated RTF-phase obtained by VAE-SSL by considering the time delay corresponding from the corresponding RTFs.

The free-space propagation delay for a given sensor is
\begin{align}
    A_i(\lambda_k,\theta) &= e^{j\frac{2\pi r_i}{\lambda_k}\sin{\theta}} \\
    A_i(f_k,\theta) &= e^{j\frac{2\pi f_k r_i}{c}\sin{\theta}}
\end{align}
with $r_i$ the sensor spacing. We assume sensor 1 the reference (with $r_1=0$) and sensor 2 the negative displacement $-r_2$. Thus the free-space RTF per \eqref{eq:mics_rtf_fft2} is
\begin{align}
    H(f_k,\theta) &= \frac{A_2(f_k,\theta)}{A_1(f_k,\theta)}\\ 
    &= e^{-j\frac{2\pi f_k r_2}{c}\sin{\theta}},
    \label{eq:synth_rtf}
\end{align}
%We thus have $\mathbf{h}=-j\frac{2\pi \mathf{F r_2}{c}\sin{\theta}$
% Taking the inverse Fourier transform of the RTF, we obtain the delta function
% \begin{align}
%     \delta(\tau) &= \int_{-\infty}^{\infty}e^{-j 2\pi k\tau} dk
% \end{align}
with $\tau=(r_2/c) \sin{\theta}$ the time delay.
% , and $k=f$ integration variable.

Given the derivation for RTF time delay in free-space, synthetic free-space RTFs are generated per \eqref{eq:synth_rtf} for the same source-microphone geometry as the DTU dataset (Sec.~\ref{sec:measurements}). The inverse fast Fourier transform (IFFT) of the RTFs is obtained, Fig.~\ref{fig:rtf_ifft1}(a). These time delays are compared to those from the IFFT of the VAE-generated RTF mean values from the reverberant DTU dataset (Sec.~\ref{sec:measurements}). In the previous section, the DTU RTF phase was reconstructed based on input the VAE. The results are shown in Fig.~\ref{fig:rtf_gen1}. One generated RTF from each of the 19 DOAs from the dataset (corresponding to phases in Fig.~\ref{fig:rtf_gen1}) were used to obtain time delays, and these delays are plotted in Fig.~\ref{fig:rtf_ifft1}(b). The maximum value of the IFFT is indicated in each plot.

Overall, the peak of the IFFT of the generated reverberant RTF means correspond to the free-space RTF delay. In Fig.~\ref{fig:rtf_ifft1}(b) the peak correlates well with the hypothesized DOA location based on $\tau=(r_2/c) \sin{\theta}$, shown as a black line in both subfigures.

We also consider the time delays corresponding to the conditionally generated RTF phase (see Fig.~\ref{fig:rtf_gen2} and \ref{fig:rtf_gen3}) from VAE-SSL trained on DTU dataset. The IFFT of the RTFs corresponding to the conditionally generated phases are shown in Fig.~\ref{fig:rtf_ifft2}. Again, the peak correlates well with the hypothesized time delay.

\begin{figure}[t]
\hspace{-4ex}
\includegraphics[width = 3.8in]{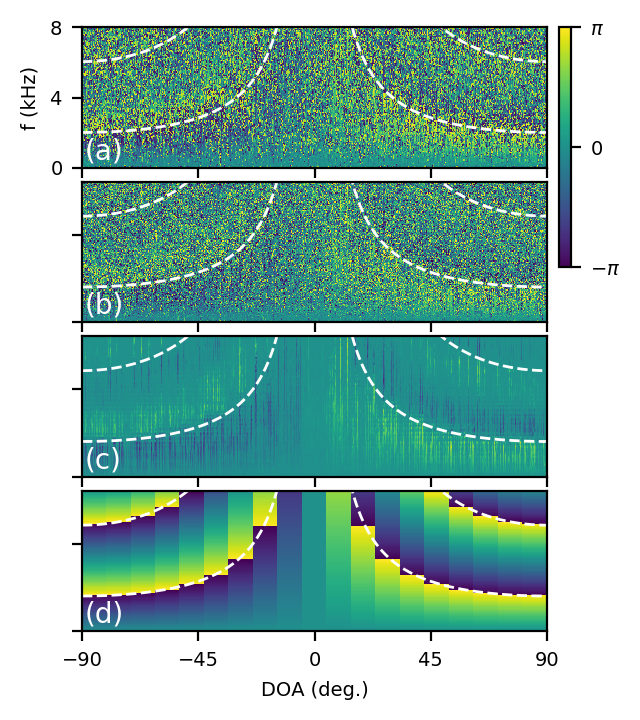}
\caption{Reconstruction of RTF phase sequences using VAE-SSL (trained using $J=988$ labels) with normal conditional distribution. Same arrangement as Fig.~\ref{fig:rtf_gen1}}
\label{fig:rtf_gen_norm}
\end{figure}

\begin{figure}[t]
% \hspace{-4ex}
\centering
\includegraphics[width = 3.0in]{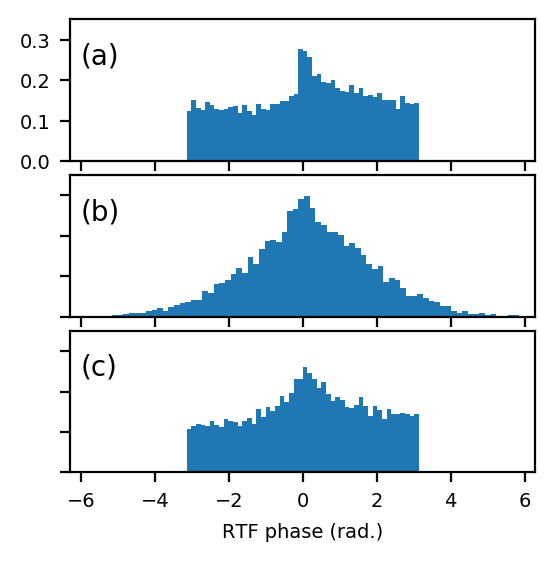}
% \subfloat[]{\includegraphics[width = 3.5in]{figures/input_dist2.png}} \\
% \subfloat[]{\includegraphics[width = 3.5in]{figures/output_dist2.png}} \\
% \subfloat[]{\includegraphics[width = 3.5in]{figures/output_dist_tn2.png}}
\caption{Histograms of RTF-phase at DOA = $20^\circ$ for 100 RTF frames from (a) input, (b) generated with normal conditional distribution $p_\theta(\mathbf{x}|\mathbf{y,z})$, and (c) generated with truncated normal conditional distribution (Sec.~\ref{sec:theory}~D--E, Sec.~\ref{sec:exper}~D).  The truncated normal distribution (c) better approximates the input RTF phase statistics (a). Histograms are normalized.}
\label{fig:rtf_gen_hists}
\end{figure}

\begin{figure}[h]
\hspace{-2ex}
\includegraphics[width = 3.5in]{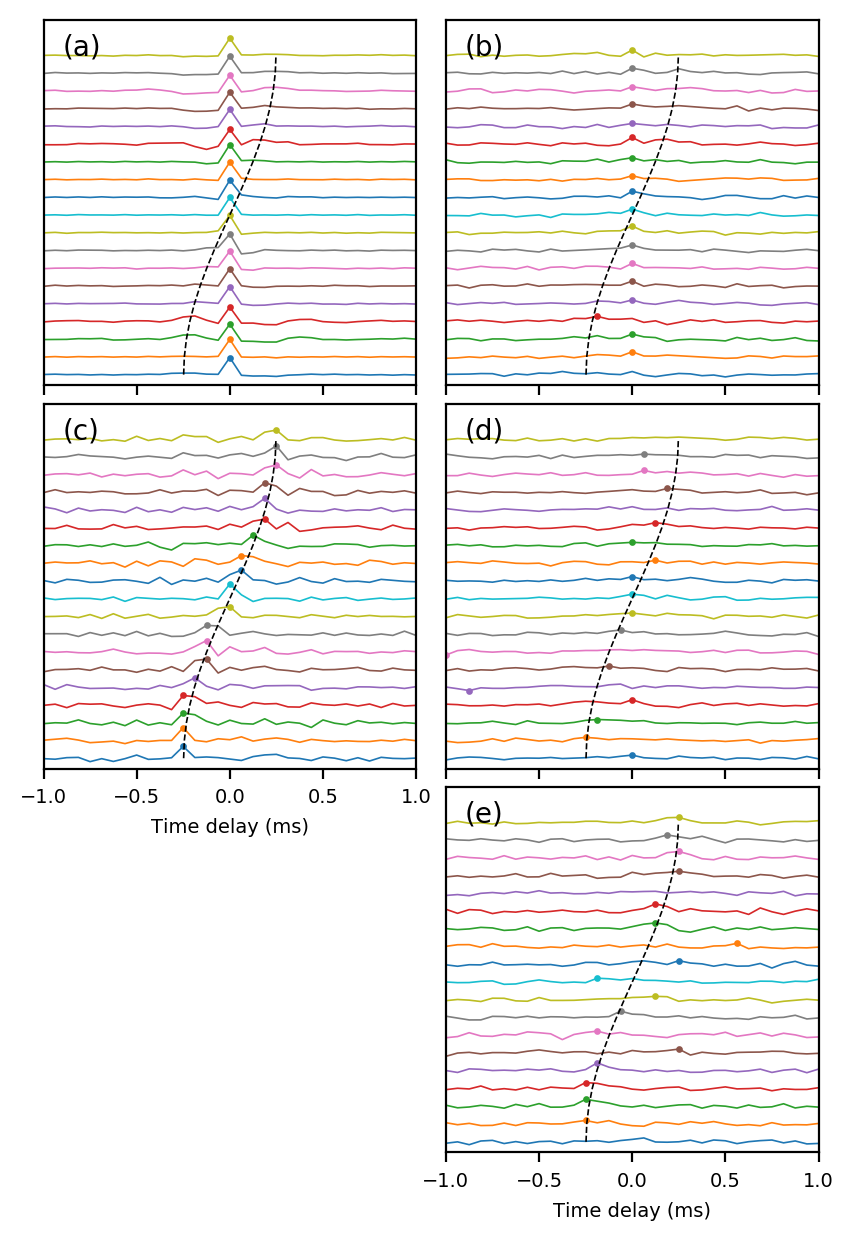}
% \centering
% \subfloat[]{\includegraphics[width = 3.5in]{figures/input_rtf.png}} \\
% \subfloat[]{\includegraphics[width = 3.5in]{figures/ifft_norm_mean_recon.png}}
% \subfloat[]{\includegraphics[width = 3.5in]{figures/ifft_norm_recon.png}} \\
% \subfloat[]{\includegraphics[width = 3.5in]{figures/phase_dtu_recon1.png}}
% \subfloat[]{\includegraphics[width = 3.5in]{figures/ifft_tn_recon.png}}
\caption{We here compare the IFFTs (time delays) of the generated RTFs using (a,b) the normal and (c,d) truncated normal conditional distribution $p_\theta(\mathbf{x}|\mathbf{y,z})$, from the VAE-SSL generative model trained on the DTU dataset. In (e) the time delays of RTFs corresponding to the input are shown. See Fig.~\ref{fig:rtf_gen1} for input and generated RTF-phase using the truncated normal conditional distribution $p_\theta(\mathbf{x}|\mathbf{y,z})$, and Fig.~\ref{fig:rtf_gen_norm} for those generated with the normal distribution. The time delays obtained using the normal distribution are biased, but with the truncated normal they are not. These results were calculated for one RTF realization per DOA ($T=19$ DOAs). The hypothesized time delay shown as black dashed line.}
\label{fig:ifft_big}
\end{figure}

As a final comparison, we consider similar analyses using VAE-SSL trained using the conventional normal distribution for the generative model conditional distribution $p_\theta(\mathbf{x}|\mathbf{y,z})$. We give results for $J=988$ labels. The results are shown in Fig.~\ref{fig:rtf_gen_norm}--\ref{fig:ifft_big}. In Fig.~\ref{fig:rtf_gen_norm}, we reconstruct the input RTF-phase sequences, similar to Fig.~\ref{fig:rtf_gen1}. In Fig.~\ref{fig:rtf_gen_hists} we compare the RTF-phase distributions generated using the normal and truncated normal distributions. In Fig.~\ref{fig:ifft_big}, we show the RTF-phase and mean RTF-phase generated by VAE-SSL using the normal and truncated normal distributions, similar to Fig.~\ref{fig:rtf_ifft1}. Since the wrapped RTF phase is on the interval $[-\pi,\pi]$, the truncated distribution to the same range is more physically meaningful. Further, the large peak of the full normal distribution-generated RTF phase (Fig.~\ref{fig:rtf_gen_hists}(b)) is related to the bias in the IFFT of the RTF corresponding to the generated phase (see Fig.~\ref{fig:ifft_big}) since zero RTF phase implies broadside source $D_{t}=0^\circ{}$. In all cases, the physical characteristics of the wrapped RTF-phase are better modeled and interpreted using the truncated normal conditional distribution. 

\section{Conclusions}

We have proposed a semi-supervised approach to acoustic source localization in reverberant environments based on deep generative modeling with VAEs, which we deem VAE-SSL. In this study, we present the first application of deep generative modeling in source localization and acoustic propagation modeling. We find that VAE-SSL can outperform both conventional signal processing-based approaches SRP-PHAT and MUSIC, as well as fully-supervised CNNs in label-limited scenarios. Further, it is shown that VAE-SSL generalizes well to source locations that were not in the training data. We demonstrate this performance using real IRs, obtained from two different environments, and speech signals from the LibriSpeech corpus. By learning to generate RTF-phase from minimally pre-processed input data, VAE-SSL models, end-to-end, the reverberant acoustic environment and exploits the structure in the unlabeled data to improve localization performance over what can be achieved simply using the available labels. In learning to perform these tasks, the VAE-SSL explicitly learns to separate the physical causes of the RTF-phase (i.e. source location) from signal characteristics such as noise and speech activity which are not salient to the task. We further showed that the trained VAE-SSL system can be used to generate new RTF-phase samples. We interpreted the generated RTF phase and verified the VAE-SSL approach well-learns the physics of the acoustic environment. The generative modeling used in VAE-SSL provides interpretable features. 

We thus observe that deep generative modeling can improve ML model interpretability in the context of acoustics. Such models are robust to light preprocessing of input features and can automatically obtain the appropriate task-specific representation in an end-to-end fashion. In our future work, we will extend this approach to multi-source and moving-source scenarios. We will further consider the effect of acoustic feature processing on the VAE-SSL generalization and sample complexity.

% \newpage

\appendices

\section{ detailed derivation of unsupervised objective} \label{app:unsup}

We here give additional details of the derivation of the objective for unlabeled data \eqref{eq:label_k4}. The steps also clarify the derivation of the supervised objective, which follows a very similar development. Starting with Bayes' rule we have for unlabeled data
\begin{align}\label{eq:bayes3_app}
p(\mathbf{y},\mathbf{z}|\mathbf{x})=\frac{p_\theta(\mathbf{x}|\mathbf{y},\mathbf{z})p(\mathbf{y},\mathbf{z})}{p(\mathbf{x})}.
\end{align}

% Direct estimation of the posterior $p(\mathbf{y,z}|\mathbf{x})$ is nearly always intractable due to $p(\mathbf{x})=\int\int p(\mathbf{x},\mathbf{y},\mathbf{z})d\mathbf{y}d\mathbf{z}$. $\theta$ indicates the parameters of the NNs used to parameterize the distributions of the generative model.

% VAEs\cite{kingma2014semi,kingma2019introduction} approximate posterior distributions using variational inference (VI)\cite{blei2017variational}, a family of methods for approximating conditional densities which relies on optimization instead of (MCMC) sampling. In VAEs the conditional densities are modeled with NNs. A variational approximation to the intractable posterior $p(\mathbf{y,z}|\mathbf{x})$ is defined by the encoder networks (two encoder networks, corresponding to $\mathbf{y}$ and $\mathbf{z}$) as $q_\phi(\mathbf{y,z}|\mathbf{x})\approx p(\mathbf{y,z}|\mathbf{x})$, with $\phi$ indicating distributions and functions defined using the encoder network. The networks constituting the VAE-SSL model are shown in Fig.~\ref{fig:ssvvae_dgm}.

Per VI we seek an approximate density $q_\Phi(\mathbf{y,z}|\mathbf{x})\approx p(\mathbf{y,z}|\mathbf{x})$ which minimizes the KL-divergence
\begin{align}\label{eq:kl_app0}
\{\Phi,\theta\}=\underset{\Phi,\theta}{\arg\min}~\mathrm{KL}(q_\Phi(\mathbf{y,z|x})||p(\mathbf{y,z|x})).
\end{align}
The KL divergence is for two arbitrary distributions is (and can be factored as)
\begin{align}\label{eq:kl_app1}
\mathrm{KL}&(Q(x)||P(x))= \int_{-\infty}^{\infty}Q(x)\log\frac{Q(x)}{P(x)}dx \\ 
&= \int_{-\infty}^{\infty}Q(x)\log Q(x)dx - \int_{-\infty}^{\infty}Q(x)\log P(x)dx \nonumber \\ 
&= \mathbb{E}_{Q(x)}\big[\log Q(x)]-\mathbb{E}_{Q(x)}[\log P(x)\big], \nonumber
\end{align}
with $\mathbb{E}_{Q(x)}$ the expectation with respect to $Q(x)$.

With the definitions in \eqref{eq:bayes3_app} and \eqref{eq:kl_app1}, the $\mathrm{KL}$-divergence \eqref{eq:kl_app0} is assessed
\begin{align}\label{eq:kl_app3}
\mathrm{KL}&(q_\Phi(\mathbf{y,z}|\mathbf{x})||p(\mathbf{y,z}|\mathbf{x})) \nonumber \\
&=\mathbb{E}\big[\log q_\Phi(\mathbf{y,z}|\mathbf{x})\big]-\mathbb{E}\big[\log \frac{p_\theta(\mathbf{x}|\mathbf{y,z})p(\mathbf{y,z})}{p(\mathbf{x})}\big] \nonumber \\
&=\mathbb{E}\big[\log q_\Phi(\mathbf{y,z}|\mathbf{x})\big]-\mathbb{E}\big[\log p_\theta(\mathbf{x}|\mathbf{y,z})p(\mathbf{y,z})\big] \nonumber \\
&\hspace{5ex} + \mathbb{E}\big[\log p(\mathbf{x})\big] \nonumber \\
&=-\mathrm{ELBO}+ \log p(\mathbf{x}),
\end{align}
with the expectation $\mathbb{E}$ relative to $q_\Phi(\mathbf{y,z}|\mathbf{x})$. 

Considering the ELBO terms from \eqref{eq:kl_app3}, we find the objective (negative ELBO) for the unlabeled data as
\begin{gather}
\begin{aligned}
-D(\theta,\phi;\mathbf{x})&=\mathbb{E}\big[\log p_\theta(\mathbf{x}|\mathbf{y},\mathbf{z})p(\mathbf{y},\mathbf{z})-\log q_\Phi(\mathbf{y},\mathbf{z}|\mathbf{x})\big] \\
&=\mathbb{E}\big[\log p_\theta(\mathbf{x}|\mathbf{y},\mathbf{z})+\log p(\mathbf{y})+\log p(\mathbf{z})\\
& \hspace{5ex} -\log q_{\phi_1}(\mathbf{z}|\mathbf{x},\mathbf{y}) -\log q_{\phi_2}(\mathbf{y}|\mathbf{x})\big].
\end{aligned} \raisetag{0.8\baselineskip}
\label{eq:elbo_app0}
\end{gather}
We can factorize the expectation in \eqref{eq:elbo_app0} with $\mathbf{y}$ and $\mathbf{z}$ independent. Since $\mathbf{y}$ has discrete states, we have for an arbitrary density $P(\mathbf{y,z})$
\begin{align} \label{eq:fact_expect}
    &\mathbb{E}_{q_{\Phi}(\mathbf{y,z}|\mathbf{x})}\big[P(\mathbf{y,z})\big]=\int\int q_\Phi(\mathbf{y,z}|\mathbf{x}) P(\mathbf{y,z}) d\mathbf{y}d\mathbf{z} \nonumber \\
    & \hspace{5ex} =\int q_{\phi_2}(\mathbf{y}|\mathbf{x}) \Bigg[ \int q_{\phi_1}(\mathbf{z}|\mathbf{x},\mathbf{y}) P(\mathbf{y,z}) d\mathbf{z}\Bigg ] d\mathbf{y} \nonumber \\
    & \hspace{5ex} =\sum_\mathbf{y} q_{\phi_2}(\mathbf{y}|\mathbf{x}) \Bigg[ \int q_{\phi_1}(\mathbf{z}|\mathbf{x},\mathbf{y}) P(\mathbf{y,z}) d\mathbf{z}\Bigg ]. 
\end{align}
Using \eqref{eq:fact_expect}, we expand \eqref{eq:elbo_app0}
\begin{gather}
\begin{aligned}
-&D(\theta,\Phi;\mathbf{x})= \mathbb{E}_{q_{\phi_2}(\mathbf{y}|\mathbf{x})}\big[ \mathbb{E}_{q_{\phi_1}(\mathbf{z}|\mathbf{x},\mathbf{y})} \big[ \log p_\theta(\mathbf{x}|\mathbf{y},\mathbf{z}) \\ 
& \hspace{10ex} +\log p(\mathbf{y})+\log p(\mathbf{z})-\log q_{\phi_1}(\mathbf{z}|\mathbf{x},\mathbf{y}) \\
&\hspace{10ex}  -\log q_{\phi_2}(\mathbf{y}|\mathbf{x})\big]\big]\\
& =\sum_\mathbf{y} q_{\phi_2}(\mathbf{y}|\mathbf{x})\big[-C(\theta,\phi;\mathbf{x,y})-\log q_{\phi_2}(\mathbf{y}|\mathbf{x})\big]
\end{aligned} \raisetag{1.4\baselineskip}
\label{eq:elbo_app1}
\end{gather}
%

% %
% \begin{gather}
% \begin{aligned}
% -&D(\theta,\Phi;\mathbf{x})= \mathbb{E}_{q_{\phi_2}(\mathbf{y}|\mathbf{x})}\big[ \mathbb{E}_{q_{\phi_1}(\mathbf{z}|\mathbf{x},\mathbf{y})} \big[ \log p_\theta(\mathbf{x}|\mathbf{y},\mathbf{z}) \\ 
% & \hspace{10ex} +\log p(\mathbf{y})+\log p(\mathbf{z})-\log q_{\phi_1}(\mathbf{z}|\mathbf{x},\mathbf{y}) \\
% &\hspace{10ex}  -\log q_{\phi_2}(\mathbf{y}|\mathbf{x})\big]\big]\\
% & =\sum_\mathbf{y} q_{\phi_2}(\mathbf{y}|\mathbf{x})\big[-C(\theta,\phi;\mathbf{x,y})-\log q_{\phi_2}(\mathbf{y}|\mathbf{x})\big]
% \end{aligned} \raisetag{1.4\baselineskip}
% \label{eq:label_k4}
% \end{gather}
% %

Thus it follows from the derivation with Bayes' rule that in the unlabeled case, we marginalize over the states of $\mathbf{y}$. This requires sampling the terms in \eqref{eq:elbo_app1} over all states of $\mathbf{y}$ for unlabeled data $\{\mathbf{x}_u\}\sim\mathcal{D}_u$.

% \newpage

\small
%\bibliographystyle{IEEEbib}
%\bibliography{biblio_gen_doa0.bib}

\EOD

\end{document}